\documentclass{emulateapj}

\usepackage{graphicx}

\newcommand{\lsim}{\mbox{$_<\atop^{\sim}$}}
\newcommand{\gsim}{\mbox{$_>\atop^{\sim}$}}

\newcommand{\arcs}{$^{\prime\prime}$}

\newcommand{\hst}{\textsl{HST}}

\slugcomment{Accepted for publication in The Astrophysical Journal}

\shorttitle{CFHTLS-Deep: Galaxy Interaction Fraction at $z<1.2$}
\shortauthors{Bridge, Carlberg \& Sullivan}

\begin{document}

\title{The CFHTLS Deep Catalog of Interacting Galaxies I. Merger Rate Evolution to $z=1.2$.}

\author{
C.R.\ Bridge,$\!$\altaffilmark{1,2}
R.G.\ Carlberg,$\!$\altaffilmark{2}
M.\ Sullivan\,$\!$\altaffilmark{3,2}}

\email{bridge@astro.caltech.edu}
\altaffiltext{1}{California Institute of Technology, Pasadena, CA 91125}
\altaffiltext{2}{University of Toronto, 50 St. George Street, Toronto,ON, M5S 3H4, Canada}
\altaffiltext{3}{Department of Astrophysics, University of Oxford,
  Keble Road, Oxford OX1 3RH, UK}

\begin{abstract}
We present the rest-frame optical galaxy merger fraction between
$0.2<z<1.2$, as a function of stellar mass and optical luminosity, as
observed by the Canada-France-Hawaii Telescope Legacy
Deep Survey (CFHTLS-Deep).  We developed a new classification scheme to
identify major galaxy-galaxy mergers based on the presence of tidal tails and
bridges. These morphological features are signposts of recent and
ongoing merger activity.  Through the visual classification of all
galaxies, down to $i_{vega}\le22.2$ ($\approx$27,000 galaxies)
over 2 square degrees, we have compiled the CFHTLS Deep Catalog of
Interacting Galaxies, with $\approx$ 1600 merging galaxies. We find the merger fraction to be
$4.3\%\pm0.3\%$ at $z\sim0.3$ and $19.0\%\pm2.5\%$ at $z\sim1$,
implying evolution of the merger fraction going as $(1+z)^m$, with
$m=2.25\pm0.24$.  This result is inconsistent with a mild or non-evolving ($m<1.5$) scenario at a
$\gsim4\sigma$ level of confidence.  A mild trend, where massive galaxies
with $M_{*}>10^{10.7}M_{\odot}$ are undergoing fewer mergers than less
massive systems ($M_{*}\sim10^{10}M_{\odot}$), consistent with the
expectations of galaxy assembly downsizing is observed.  Our results also show that interacting galaxies have on average
SFRs double that found in non-interacting field galaxies. 
We conclude that
(1) the optical galaxy merger fraction does evolve with redshift, (2)
the merger fraction depends mildly on stellar mass, with lower mass
galaxies having higher merger fractions at $z<1$, and (3) star formation is triggered at all
phases of a merger, with larger enhancements
at later stages, consistent with $N$-body simulations. 

\end{abstract}
\vspace{0.2cm}
\keywords{galaxies: evolution -- galaxies: formation -- galaxies:interactions -- galaxies: starburst}

\section{Introduction}
In the past three decades the concept of galaxies as ``island universes''
slowly evolving in isolation has changed dramatically.
Gravitational interaction between galaxies is now considered a relevant
factor in a galaxy's evolution, capable of altering its morphology,
luminosity, color, size, star formation rate (SFR), and mass distribution, all over a
relatively short timescale.  According to the $\Lambda$-dominated cold
dark matter ($\Lambda$CDM) model, the merger rate of dark matter halos and
similarly the merger rate of galaxies are the most fundamental
processes in structure formation. Numerous $N$-body simulations and
semi-analytical models have studied the merger rate of dark matter
halos, predicting evolution with redshift as $(1+z)^{m}$, with
$1.0<m<3.5$
\citep{2001ApJ...546..223G,2006ApJ...652...56B,2008MNRAS.386..577F}.  
A direct comparison between dark matter halo merger rates
and galaxy merger rates is difficult due to the uncertainty in the galaxy halo occupation
number.  Although measuring the frequency that galaxies merge as a function of
cosmic epoch can place powerful constraints on the theory of galaxy
evolution and structure formation.

Traditional observational approaches aimed at investigating the major galaxy merger
rate, measure the frequency of galaxy mergers or galaxies in close
pairs, spanning a range of redshifts.  The merger or close pair fraction
should evolve in a similar manner as the merger rate assuming the
timescale over which the merger selection criteria are sensitive to
is independent of redshift. \citet{1977egsp.conf..401T} was the first to suggest that the merger
rate may be larger at higher redshifts, by using estimates of past
merger remnants.  Numerous studies have estimated the
evolution of the merger rate, yielding highly discrepant measurements of
$m$, ranging from no evolution ($m\sim0$) to strong evolution
($m\sim5$)
\citep[][to name a few]{1989ApJ...337...34Z,1994ApJ...435..540C,2000ApJ...532L...1C,1997ApJ...475...29P,2000MNRAS.311..565L,2003AJ....126.1183C,2004ApJ...617L...9L,2007ApJ...659..931B,2007ApJS..172..320K,2008ApJ...672..177L}
.  It is often suggested that this discrepancy stems from
the variety of techniques used to identify galaxy mergers, incompleteness
corrections, as well as differences in sample selection.  Although
these factors likely play a role, typical sample sizes have been on the
order of a few ten's of mergers (excluding
\citet{2007ApJS..172..320K}) resulting in $>30\%$ uncertainty in the
evolution of the merger rate and the effect of cosmic variance has
been largely unexplored.

Over the past decade it has become clear that the average SFR per unit comoving volume (CSFR) has declined by an
order of magnitude since $z\sim1$
\citep{1996ApJ...460L...1L,1998ApJ...498..106M}. A fundamental
question that remains is the cause of the drop of the CSFR.  It has
been suggested by some close pair
\citep{1994ApJ...429L..13B,1995ApJ...445...37Y,1997ApJ...475...29P,2002ApJ...565..208P,2007ApJ...659..931B,2007ApJS..172..320K}
and merger studies
\citep{2000MNRAS.311..565L,2003AJ....126.1183C,2007ApJS..172..329K}
that an increased merger rate at higher redshift is either partially
or completely responsible for the higher SFR density
at $z\sim1$.  However, there are merger studies which have found
little or no evolution of the merger rate
\citep{2000ApJ...532L...1C,2004ApJ...601L.123B,2004ApJ...617L...9L,2008ApJ...672..177L},
indicating the decrease in the volume-averaged SFR
density since $\sim1$ is not a result of a declining merger rate but rather
declining star formation in disk galaxies \citep{2008ApJ...672..177L}. 

In order to probe the merger rate a clear definition of what
constitutes a merger and a robust identification technique are required.  One traditional method is
to search for close galaxy pairs.  However, even with radial velocity
measurements for both galaxies, about half of all physical pairs may be
chance superpositions \citep{2000ApJ...536..153P,2002ApJ...565..208P}.    Another method is to calculate internal asymmetries,  but these quantitative
morphological parameters are subject to complication, since star
formation itself is sufficiently violent and chaotic and therefore not
smoothly distributed, potentially mimicking features of a merger when
none has occurred.  However, long tidal tails are nearly a foolproof
signature of imminent mergers.  Moreover, tidal tails are a relatively
simple, completely dynamical phenomenon that can be studied in detail
with $N$-body simulations \citep{1972ApJ...178..623T,1992ApJ...393..484B,2008ApJS..175..356H}.  In spite of these positive features little
work on galaxies with tidal tails has been done at high redshift. The
reason is simple: tails have a relatively low surface brightness.  

The Deep component of the Canada France Hawaii Legacy Survey
(hereafter CFHTLS-Deep)
provides some unique advantages for studying morphologies
and the merger rate of galaxies.  The deep ground--based
imaging ($i'<27.3$) is ideal for the detection of low surface brightness
features like tidal tails.  The survey area is also spread over four fields allowing for cosmic
variance estimates of the merger rate, while the larger area allows for
merger fraction estimates no longer dominated by small
number statistics as seen in previous studies \citep[excluding][]{2007ApJS..172..320K}.

In this paper, we apply a new technique for identifying major merger
candidates in the CFHTLS-Deep Survey.  Through the visual classification of $\approx$27,000
galaxies over 2 square degrees, we have complied the {\bf \it CFHTLS-Deep Catalog of Interacting
  Galaxies}, which contains $\approx$1600 galaxies
between $0.1<z<1.2$ with tidal tails and bridges.  This is the largest
catalog of its kind in the literature.  In Section 2, we describe the
optical photometry from the CFHTLS-Deep Survey, along with the methodology used for deriving the photometric
redshifts, stellar masses and SFRs.  Section  3
outlines the technique used to identify interacting galaxies,
including completeness tests.  We measure the galaxy interaction
fraction in Section 4, its
dependence on optical luminosity, stellar mass, and address cosmic
variance.  With additional assumptions about the merger timescale we estimate the merger rate and
interaction history of galaxies from $z\sim1$ in Section 5, followed by the impact
mergers have on the SFRs of galaxies (Section 6).  We summarize our
conclusions in Section 7. 

In the discussion that follows, we assume Vega magnitudes and any calculation requiring cosmology assumes $\Omega_{\rm M}=0.3, \Omega_\Lambda=0.70$, and H$_0$=70\,km\,s$^{-1}$\,Mpc$^{-1}$.  

\section{The CFHTLS Deep Observations}
\subsection{MegaCam Optical Imaging}
The CFHT Legacy Survey (CFHTLS) is a joint community project between Canada
and France with more than 450 nights over a 5 year period that
commenced in 2003 June. The survey exploits the square degree
MegaCam camera \citep{2003SPIE.4841...72B} on CFHT which has 36, 2048$\times$4612
pixel CCDs with a pixel scale of 0\arcs.187.  The Deep survey is one component of the CFHTLS (deep, wide, very wide), covering four low
Galactic extinction fields in $u^{*}, g', r', i', z'$.  Each survey
field, named D1 through D4 is one
square degree and distributed in right ascension (R.A.) for efficient observing
throughout the year.  The survey fields were centered at J2000 R.A.$=
02^{h}26^{m}00^{s}$, decl.$=-04^{\deg}30^{'}00^{''}$ (D1), R.A.$=
10^{h}00^{m}29^{s}$, decl.$=+02^{\deg}12^{'}21^{''}$ (D2), R.A.$=
14^{h}17^{m}54^{s}$, decl.$=+52^{\deg}30^{'}31^{''}$ (D3), R.A.$=
22^{h}15^{m}31^{s}$, decl.$=-17^{\deg}44^{'}05^{''}$ (D4).  The
analysis that follows utilizes only the D1 and D2 fields, a total of 2 square degrees, due to their
extensive ancillary data \citep[see][for details]{2006AJ....131..960S}.

\begin{deluxetable}{cccc}
\tablewidth{250.1pt}
\tabletypesize{\footnotesize}
\tablecaption{\label{tab:exptime} CFHTLS-Deep: Final Stacks}

\tablehead{\colhead{Field} & 
\colhead{Filter} & 
\colhead{Int. Time [hrs]} &
\colhead{Limiting Magnitude}}

\startdata
\tableline
D1 ...........  & u* & 10.6h & 26.7\\
(0.96 square degree)         & g' & 9.5h &27.1\\
         & r' & 18.8h & 26.8\\
         & i' & 45.1h & 26.4\\
         & z' & 20.0h & 25.6\\
\hline
D2 ...........  & u* & 3.4h & 26.4\\
(0.90 square degree)        & g' & 5.7h & 27.0\\
         & r' & 10.7h & 26.5\\
         & i' & 22.2h & 26.2\\
         & z' & 12.0h & 25.2\\
\enddata
\tablecomments{Limiting magnitudes are estimated by adding artificial galaxies, faded and rescaled into the images and then trying to recover them.}

\end{deluxetable}

The optical images used to derive the galaxy parameters, and
morphological classification were constructed
by the Supernova Legacy Survey \citep[SNLS;][]{2006ApJ...648..868S}, a
key project in the CFHTLS.  Deep optical stacks were generated for
each filter ($u^{*}, g', r', i', z'$) with an imposed wavelength-dependent seeing limit ($\sim$4 pixels) to
ensure a high degree of resolution.  The goal was to maximize exposure
depth while retaining similar and excellent seeing in the different
filters.   The typical seeing of the final
stacks is 0\arcs.7-0\arcs.8 ($i'$ band). The individual ``Elixir''
processed images are available from the Canadian Astronomy Data Centre
(CADC); however, the final stacks and photometric catalogs including
redshifts in which this work is based are not currently public.  The details of the processing technique are discussed in
\citep{2006ApJ...648..868S}.

Source extraction and photometry were performed on each Deep field using
SExtractor \citep{1996A&AS..117..393B} in dual image mode. Detections
were performed in the $i'$ filter ($i'\sim26.3$)
and photometry measurements calculated in each of the five filters,
$u^{*}, g', r', i', z'$.  The exposure times and limiting magnitudes
reached for each filter in both fields are given in Table
\ref{tab:exptime}.  We applied a bright star and bad pixel mask to the images prior to running the source detection to
eliminate noisy or contaminated regions.   The total area masked is
$\le10\%$ for each field.  We compared our number counts to those from
the Hubble Deep Field (HDF) North and South
\citep{1996AJ....112.1335W,2001MNRAS.323..795M} and COSMOS
\citep{2007ApJS..172..219L}. They were found to be consistent for
galaxies brighter than $i'<24$, on which this work is focused.

\subsection{Galaxy Properties: Redshift, Stellar Mass, SFRs}
\label{sec:galaxy-prop}

In this section, we outline the methods used to convert the optical
fluxes of our sources into a photometric redshift, and derive other
properties such as stellar mass and SFR.

In order to study the potential evolution of the interaction and star formation rates of galaxies, we need to derive a redshift estimate for each galaxy.  Although spectroscopic redshifts are the most precise
distance measures, they are observationally expensive to obtain for large
samples.  A strength of the CFHTLS-Deep survey is its high quality
five-band optical imaging, which can be used to derive a photometric
redshift estimate.  The five optical bands can be combined to produce a broadband
spectral energy distribution (SED) that can be compared to a set of
template SEDs to estimate properties of the galaxy such as its redshift, age, stellar mass,
and SFR \citep{1962IAUS...15..390B,1986ApJ...303..154L}.  This
technique uses broad spectral features like the 4000\AA\, discontinuity
or the Lyman break for comparison to templates.  

We fit a series of galaxy template SEDs to the broadband fluxes of each
galaxy.  The best-fit SED is determined through a standard $\chi^{2}$
minimization procedure between the synthetic photometry generated by integrating
the template SEDs through the CFHTLS filters, and the observed fluxes
(including the flux errors). This was performed with the Z-Peg template fitting
code \citep{2000A&A...363..476B,2002A&A...386..446L}.  Its methodology is similar to that applied
by other photometric redshift codes \citep[e.g.][]{1996ApJ...468L..77G}.

We employ a set of synthetic templates computed with the
\textsc{PEGASE-ii} galaxy evolution code
\citep{1997A&A...326..950F,2002A&A...386..446L,2004A&A...425..881L}.
Both the SEDs and the photometric redshift code have been extensively
tested and used in the literature \citep[e.g.][]{2004ApJ...614L...9M,2006A&A...449..951G,2006ApJ...648..868S}.
We use eight evolutionary scenarios that evolve self-consistently with age and
assume a Rana-Basu initial mass function \citep{1992A&A...265..499R}.   These scenarios were
designed to match the average colors of local galaxies and to
reproduce deep galaxy number counts.  Considering several
tens of time steps for each scenario, the total number of synthetic
SEDs is $\approx$500.  When fitting a given
galaxy only templates younger than the age of the universe at the
redshift of the galaxy are considered.  
The accuracy of the photometric redshifts is determined by comparing
them to the SNLS spectroscopic sample in our fields
\citep{2005ApJ...634.1190H,2008A&A...477..717B}.  The photometric redshift accuracy down to
$i'\sim22.5$ is $\sigma_{\Delta z}/(1+z_{s})=0.04$.

Two additional physical parameters of particular interest are the SFRs of
interacting galaxies and their stellar masses. These quantities are
both derived using the Z-Peg code \citep{2002A&A...386..446L}.  The
$mean$ recent SFR for a galaxy is determined from the best-fit SED, averaging
the SFR over a period of 0.5 Gyr.  The total stellar mass of a galaxy was derived by integrating the
total star formation history (SFH) of the best-fit scenario, up to the
best-fit age and subtracting off the mass from stars that have died.
The results from these techniques were found to be in
good agreement with template fits when a spectroscopic redshift was
used \citep{2006ApJ...648..868S}. The systematic
errors associated with these techniques, their application to CFHTLS data, and consistency checks are outlined in \citet{2006ApJ...648..868S}.

\section{CFHTLS-Deep Catalog of Interacting Galaxies}
\subsection{Identifying Interacting Galaxies}
\label{subsec:indentify}

The first step in studying the frequency of galaxies involved in an
interaction is to define a clean, robust, and useful definition of an
interacting galaxy.  Morphologically, interacting galaxies can exhibit
long tidal tails, bridges (linking two or more galaxies), ring structures,
stellar bars and/or enhanced spiral structure and frequently appear severely distorted.  We have focused our identification methods on
confidently selecting galaxies which have recently undergone or are presently
undergoing a tidal interaction.  The presence of a tidal tail or
bridge is incontrovertible evidence of a recent interaction. 

There are two different avenues one can take to morphologically select
interacting galaxies.  The first is to utilize quantitative
morphological software that measures a galaxy's structural
parameters, such as asymmetry, Gini or $M_{20}$
\citep{1996ApJS..107....1A,1996MNRAS.279L..47A,2000ApJ...529..886C,2004AJ....128..163L}.
As a galaxy undergoes a merger, tidal fields distort the galaxies radially, drawing out galactic
material into long plumes and tails.  These structural parameters can help
describe the level of a galaxy's disturbed appearance and are commonly
used to identify mergers.
 
This is an efficient and automated approach.  However, a complication of
this method is the need for high-resolution data, especially at
high redshifts when Hubble Space Telescope (\hst) observations are a requirement.  A further
complication is that \hst\,/ACS images do not easily detect low surface brightness features, like
tidal tails due in combination to the small plate
scale and point spread function (PSF).

The second method of identifying interacting galaxies is based on pure
visual inspection.  This technique has been utilized by many in the
past to morphologically classify galaxies
\citep[][to name a
few]{1926ApJ....64..321H,1959VV....C......0V,1961hag..book.....S,1966ApJS...14....1A}.
Qualitative classification is able to identify the low surface brightness features
that the automated method has difficulties with, as it utilizes one of
the best pattern recognition computers--the brain-eye
combination. A drawback of this method is that it can be laborious, subjective and can suffer from reproducibility issues when
the person conducting the classification changes.  However, when a set of visual criteria are clearly
defined, and the features themselves are striking (e.g., long tidal
tails and bridges), visual inspection can be a highly accurate method
of morphological classification (see Section \ref{sec:class_exp}).

The ground-based nature of our dataset coupled with its depth make it
ideal to detect low surface brightness features down to $<$29
mags arcsec$^{-2}$.  Ultimately, we define an
interacting galaxy to be one with a tidal tail or bridge.
These merger signatures are visible after a first encounter and can
persist even after the galaxies' nuclei have coalesced.  Although our primary objective was
to identify galaxies undergoing an interaction, other classifications
were included and will be presented in a future paper.

The morphology of galaxy interactions relies on a large number of
variables, such as the geometry of the encounter, impact velocity, and
mass ratios \citep{1992ApJ...393..484B}.  By defining a
sample of tidal tailed galaxies we are selecting galaxy mergers
which have recently experienced a merger of mass ratio
$M_{1}/M_{2}>1/10$.  Galaxy encounters with mass ratios $>1/10$ have
been shown to have significant impact on a galaxy's evolution
\citep{1992ApJ...393..484B,1996ApJ...464..641M}.
Although some of the tailed galaxies in our sample will have been a
result of lower mass ratio or minor mergers (defined as those with $1/10<M_{1}/M_{2}<1/4$), the majority are likely to have
been the consequence of $M_{1}/M_{2}>1/4$ interactions, due to the
marked nature of the tidal features before being included in our sample \citep[e.g.][]{1992ApJ...393..484B,1996ApJ...464..641M,1999ApJ...526..607D}.

\subsection{Classification Methodology and Scheme}
\label{sec:classification}

The large area and deep optical imaging of the CFHTLS-Deep survey
makes it well suited to identifying interacting galaxies.  All
galaxies in the D1 and D2 fields (2 square degrees) down to $i_{vega}=22.2$ ($\approx$ 27,000) were visually
classified using the survey's deepest filter, the $i'$ band.  The
apparent magnitude distributions were similar for the D1 and D2
field.  There were small differences ($<10\%$) at the bright end but were due
to the D1 field covering 0.05 square degrees more than D2, coupled
with a larger number of bright stars in the D2 field, resulting in a
larger masked out region.   

Each galaxy was viewed on a computer screen with DS9 \citep{2003ASPC..295..489J}, and the classification
was logged through keypad entry to a file. More than $90\%$ of galaxies with apparent magnitudes
brighter than $i'\le$21.9 were able to be visually classified.
Although some galaxies at these apparent magnitudes had angular sizes
too small for a reliable classification to be made, the
majority of the remaining $10\%$ lied within halos of bright stars, or
bad regions of the images that were not masked out.  At
faint magnitudes visual classification becomes increasingly difficult
and less reliable, due typically to the smaller angular size of the objects.  We therefore imposed an apparent magnitude cutoff of
$i'_{vega}\le$21.9 to ensure highly reliable classifications. Our final
morphological catalog consists of 25,194 galaxies.

Galaxies were classified as interacting or
merging if they exhibited a tidal tail or tidal bridge between
galaxies.  These two tidal signatures encompass early through to advanced stage mergers. We then further
sub-classified these objects in a simplistic way according to the length of their longest
tidal tail as well as the number of tails to
see if they were correlated with other properties such as the SFR or active galactic  nucleus (AGN) activity.  The length of the tidal tail was defined in relation to the angular
size of the host galaxy. Figure 1 provides $i'$ band and color composite ($g',r',i'$) examples of galaxies classified as interacting. Galaxy mergers were identified and classified into the following primary categories:\\

\indent{\it{Long tidal tailed galaxies.}}-- Galaxies exhibiting a tidal tail
longer than the diameter of the host galaxy.  These mergers primarily
represent intermediate to late stage interactions after a first
encounter.  \\

\indent{\it{Medium tidal tailed galaxies.}}-- These systems have a
tidal tail length approximately equal to the diameter of the host
galaxy, and also probe interactions after the first passage to advanced
mergers.\\

\indent{\it{Short tidal tailed galaxies.}}-- This classification
identifies galaxies with tidal tails that are less than the diameter
of the host galaxy.  This classification suffers from the most
contamination from spiral arms being misidentified as tidal tails.
Therefore, short tidal tailed galaxies are only included in the potentially interacting sample and not considered in the primary
analysis of this paper.\\

\indent{\it{Tidally bridged galaxies.}}-- Galaxies found to have a tidal bridge linking it to another galaxy.  This
classification represents a narrower time frame in the merger sequence,
typically immediately after the first encounter, or second passage,
and is in turn well suited to comparative studies with simulations.\\

\indent{\it{Double nuclei.}}-- Our final primary merger classification
identifies mainly advanced stage mergers, with two or more nuclei
overlapping in a common envelope.  Since some galaxies with apparent double
nuclei are a result of line-of-sight projections this
classification was further separated into those galaxies with a
multiple nuclei and additional evidence of a tidal tail and those with no
tidal tail signatures. \\

Table \ref{tab:class2} outlines the
classification scheme and the number of galaxies identified in each
merger class. Ultimately, we find 1586 galaxies with clear
signatures of a tidal tail or bridge, with an additional 970 potentially interacting galaxies.

The sample of potentially interacting galaxies is comprised of galaxies with a double nuclei with no evidence of a tidal tail, which can be a result of projection, and those with ``short'' tidal tails since these tails are less pronounced and can be confused with spiral arms.  However,  if the ``short'' tailed galaxy had secondary evidence, such as the presence of a double nuclei or a tidal bridge they were considered to be undergoing a merger.   Since an aim of this paper was to measure a lower limit of the fraction of galaxies undergoing a major merger the sample of potentially interacting galaxies was included only as secondary measure of the merger fraction (see Section \ref{sec:galaxy-inter-fract}) and not in the merger rate determination or the general conclusions of this work.

\begin{deluxetable}{ll}
\tablewidth{250.1pt}
\tabletypesize{\footnotesize}
\tablecaption{\label{tab:class2}Classification Scheme for Interacting Galaxies}

\tablehead{\colhead{\textsl{{General Description}}} & 
\colhead{\textsl{N}}}

\startdata

``Long Tidal Tails'': Tail Length $>$ Diameter of Host & \\
\hline
Galaxies with a long tidal tail: 1 tail & 157\\
Galaxies with a long tidal tail: 2 tails & 183\\
Galaxies with a long tidal tail: $>2$ tails& 36\\

\\
\hline

``Medium Tidal Tails'': Tail Length $\sim$ Diameter of Host & \\
\hline
Galaxies with a medium length tidal tail: 1 tail &293 \\
Galaxies with a medium length tidal tail: 2 tails &138\\
Galaxies with a medium length tidal tail: $>2$ tails&25\\

\\
\hline

Close Galaxy-Galaxy Pairs with Tidal Bridges & \\
\hline
Galaxy in a close pair with bridge, no tail &486\\
Galaxy in a close pair with bridge, + short tidal tail &48\\
Close Pair with bridge + medium tail(s) &49\\
Close Pair with bridge + long tail(s) &50\\
\\
\hline

Double Nuclei (DN) & \\
\hline

Galaxies w/ a DN + short tail(s)  &12\\
Galaxies w/ a DN + medium tail(s) &47\\
Galaxies w/ a DN + long tail(s) &62\\
\\
\hline

\bf Total Number of Confidently Interacting Galaxies  &1586 \\
\hline
\hline
\\

Potential Galaxy Interactions & \\
\hline
 ``Short Tidal Tails'': Tail Length $<$ Diameter of Host & 586 \\
Galaxies with a double nuclei (DN), no tidal tail & 384\\
\enddata
\tablecomments{$N$ specifies the number of galaxies in each morphological classification}
\end{deluxetable}

\begin{figure*}
\begin{center}
\includegraphics[width=6.5in]{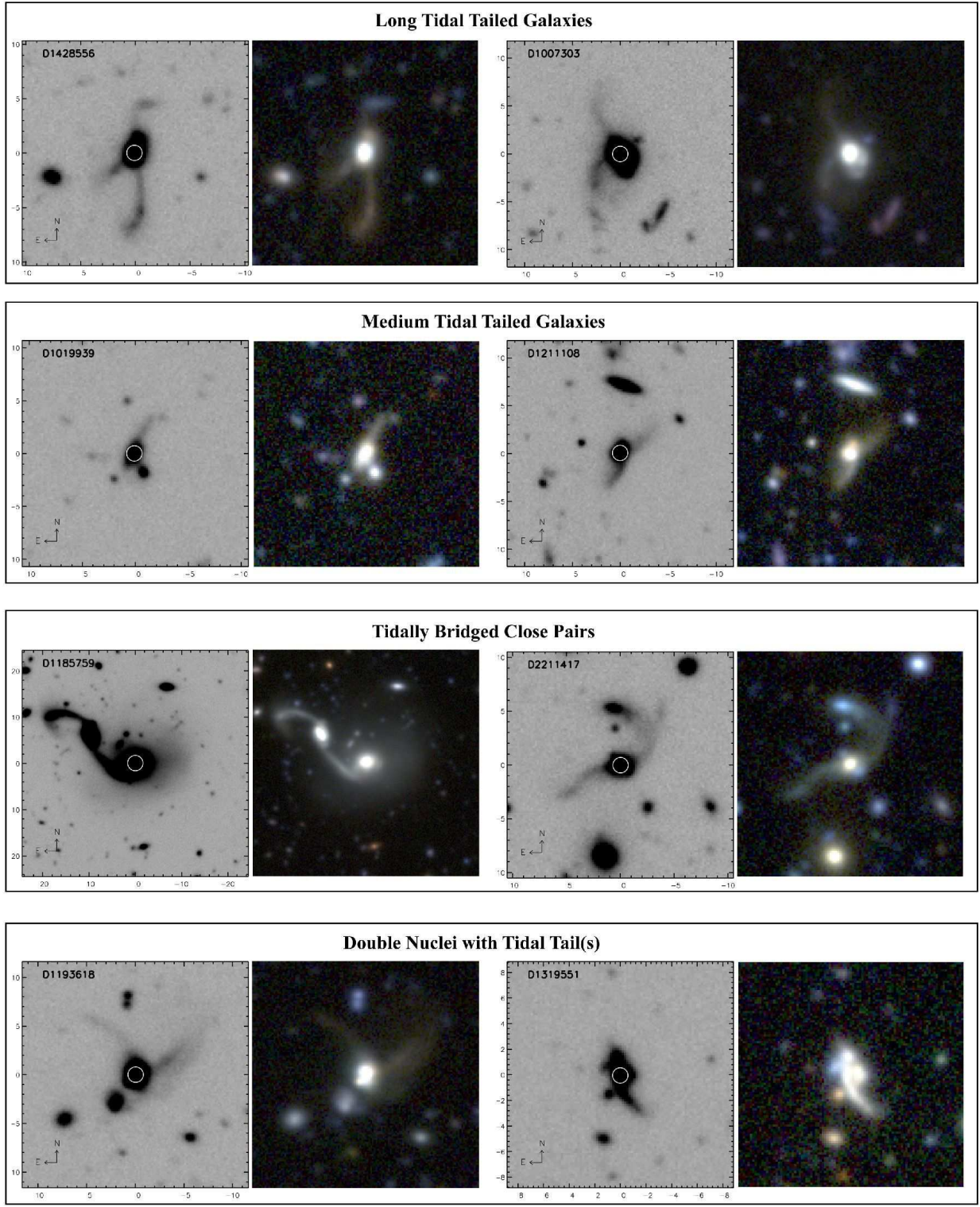}
\caption[Examples of Interacting Galaxies]{Examples of galaxies classified to have tidal tails or
bridges from the {\it CFHTLS-Deep
  Catalog of Interacting Galaxies}.  The $i'$ band images (left) and
  color composites ($g',r',i'$) range in class from galaxies with
 tidal tails (various lengths), close pairs with tidal bridges to
  double nuclei with tidal tails.  Each stamp is 100$h^{-1}$kpc on a
  side.  The white circle marks the galaxy that has been
  classified. The $XY$ axes are in arcseconds.}

\end{center}
\label{fig:stamps}
\end{figure*}

\subsection{A Re-Classification Experiment}
\label{sec:class_exp}

A potential bias in using a visual classification technique is
its reproducibility when other individuals inspect the same sample of galaxies.  A key criteria of this work, however, was the
requirement of strong tidal signatures before a
galaxy was deemed ``interacting'', dramatically reducing the classification variance by different
individuals.  C.B. visually classified all galaxies under study. To cross-check these
classifications, M.S. and K. Bundy (KB) classified a sub-sample of 700
objects. 
This sample of galaxies was randomly selected from within four
weighted parent
classifications (490 galaxies with tidal tails, 110 galaxies with a
tidal bridge, 50 galaxies classified as potentially interacting, and 50 deemed non-mergers). 

This re-classification experiment was a blind study, both K.B. and M.S. were given no information regarding the galaxies they were classifying.  The classification was
performed in the same manner as described in Section
\ref{sec:classification}. Of the galaxies used in the
classification experiment, $86\%$ had been categorized as
``interacting'' prior to the experiment (by C.B.).  Both K.B. and M.S.
classified $88\%$ and $87\%$ of the galaxies in the experiment to
be ``interacting''.  The additional $1\%-2\%$ found by K.B. and M.S. to be
interacting were predominantly classified as
``potential'' galaxy interactions with $<0.5\%$ deemed non-mergers by C.B. The strong agreement between different
individuals regarding which galaxies are
``interacting'' is a result of the robust visual criteria a galaxy
must exhibit before it is classed as undergoing an interaction.  We are therefore excluding many potentially
interacting systems as well as those with tidal
features below the sensitivities of our survey.  The results
that follow are therefore secure lower limits.

As an additional test, C.B. reclassified a set of 500 galaxies (350
interacting, 150 non-interacting) randomly selected. This blind self test addresses the reproducibility of
the authors own classifications.  The classifications remained
the same $97\%$ of the time.  The $3\%$ variation was primarily a
result of galaxies being classed into morphologically similar types,
for example, an intermediate tidal tailed galaxy being classed as having a short
tidal tail, and not into grossly different types.

\subsection{Classification Completeness}
\label{sec:recovery}

When investigating how the fraction of interacting
galaxies changes from low-$z$ to high-$z$, it is important to
investigate how tidal features become less resolved and fainter due to
cosmological effects.  To address this question we have artificially
redshifted bright nearby galaxies with tidal features in our sample
out to higher redshifts.  We then reclassify the redshifted galaxies
to determine the point in redshift space that the tidal signatures are
too weak, and the galaxy would not be classified as interacting. 

When simulating an image of a low redshift galaxy to how it
would appear at a higher redshift, multiple factors must be
considered.  First, we rebin the image, ensuring that flux is
conserved, to account for the reduction in apparent size of the galaxy
when it is viewed at higher redshift.  Second, the dependence of surface
brightness ($SB$) with redshift (SB$\propto(1+z)^{-4}$), as well as
$k$-correction effects needs to be accounted for.  In order to
accurately simulate the appearances of galaxies at high redshift we
carefully selected the lowest redshift galaxies possible with similar luminosities to those probed at
the high redshift end of our data ($z\sim0.7-1.0$).  Our resultant sample
consisted of 54 galaxies with tidal features between $0.3\le z\le0.45$
and $M_{g}\le-21.0$. 

At redshifts between $0.4\le z \le 0.9$, the $i'$-band, which was the
filter used for visual classification, is
probing rest frame $g'$. Therefore, to help account for the
$k$-correction the $r'$-band images of our sample of low-$z$ galaxies ($0.3\le z\le0.45$) 
were artificially redshifted, as they also probe rest $g'$, like that of the
$i'$-band images used for classification of the 
main sample.  When redshifting the low-$z$ galaxies to our highest redshift bin ($z\sim$1-1.15),
ideally the $g'$-band images should be used so that the same rest-wavelengths are being compared in original and redshifted samples.  However, the $g'$ images were not sufficiently deep, and in turn the $r'$-band images were used.

Ultimately, a sample of 54 low redshift galaxies were redshifted out to five
different redshifts, $z=$0.55, 0.70, 0.85, 1.0 and 1.15.  The galaxies were
re-classified at each interval for the presence of tidal features.  The recovery rate remains high out to $z\sim0.85$, where we can still
identify strong tidal features in $\sim80\%$ of the redshifted
galaxies.  Throughout the analysis, the recovery rate ($75\%$ at $z=0.9$) is referred to as a
completeness limit, as it quantifies our survey's sensitivity to tidal
features. The rate of recovery could also be used as a correction
factor when determining the fraction of galaxy mergers at a given
redshift (as discussed in Section \ref{sec:galaxy-inter-fract}).

\subsection{Sample Selection}
\label{sec:sample-selection}

In order to trace the fraction of merging galaxies with redshift, we first
need to define a sample of galaxies that probe the same stellar mass or
luminosity ranges over the entire redshift range.  In Sections
\ref{sec:classification} and \ref{sec:recovery}, we
found that visual classifications could be securely made to
$i_{vega}\le21.9$, with $>75\%$ completeness out to $z=0.9$. 

\begin{figure}
  \centering
  \includegraphics[width=85mm]{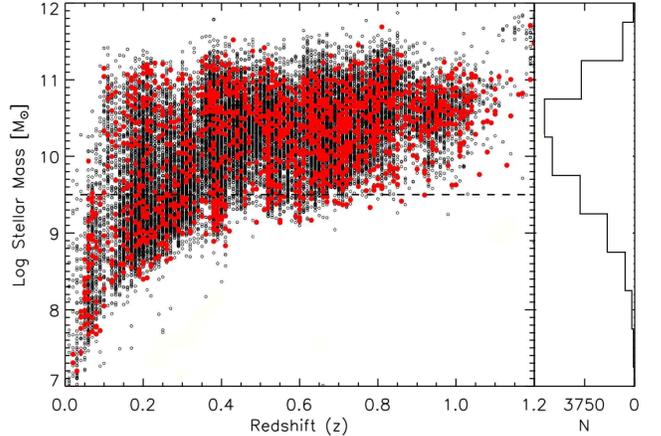}
\caption[Stellar Mass vs. Redshift: CFHTLS]{Stellar mass as a function
  of redshift for the CFHTLS D1 and D2 fields. Galaxies classified as
  non-interacting are shown as open circles (black) and those undergoing an interaction, filled
  circles (red).  The dashed line defines the stellar mass limit
  (log mass $>9.5 M_{\odot}$) imposed.  The
  histogram on the right notes the stellar mass distribution of the field.}
\label{fig:cfhtmass}
\end{figure}

The majority of merger rate studies to date impose luminosity rather
than stellar mass limits on their data sets \citep[excluding][]{2003AJ....126.1183C,2008MNRAS.386..909C,2009ApJ...697.1369B}.
A potential bias with this selection criteria is the inclusion of
interacting galaxies into the sample which prior to the merger event
may have been fainter than the luminosity cutoff.  This selection bias
may lead to higher merger fractions or close pair counts \citep{2004ApJ...601L.123B,2006ApJ...652...56B,2008ApJ...685..235P}.  However, stellar mass estimates are less affected
by merger induced star formation \citep{2006ApJ...638..686C}, and therefore a less biased selection criteria.  In the
sections to follow we explore the galaxy interaction fraction
using a stellar mass limit of $log M_{*} (M_{\odot}) \ge
9.5$ (see Figure \ref{fig:cfhtmass}), however also consider luminosities limits to fairly compare our
findings with previous works.  The stellar mass and luminosity limits were chosen by balancing incompleteness with sample size.

To first order, our sample was divided into ``interacting'' and ``non-interacting''
galaxies.  As outlined in Section \ref{subsec:indentify}, we define ``interacting'' or
``merging'' galaxies as those with tidal tail features, more
specifically, with the following visual classifications: medium to long
tidal tails, galaxies in a close pair linked with a tidal
bridge, and those with a clear double nuclei and a
tidal tail.  We find a total of 1586 interacting galaxies within the
D1 and D2 fields combined.  

In order to investigate the properties of interacting and merging
galaxies, we also need to establish a comparison sample of
non-interacting galaxies.  We combine all the ``non-interacting''
classifications, which are primarily composed of spiral or disk
dominated sources to construct a fair comparison sample.  We have
identified 22,268 non-interacting galaxies.  

\begin{figure}
  \centering
  \includegraphics[width=90mm]{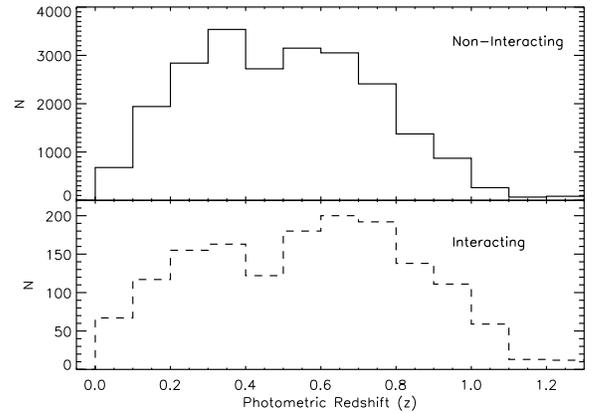}
\caption[Redshift Distribution: Interacting and Non-Interacting Galaxies]{Redshift distribution for all galaxies classified as non-interacting
  (upper) and interacting (lower, dashed). The mean photometric redshifts for the two
  samples are 0.51 (non-interacting) and 0.58 (interacting). The two
  samples have comparable redshift distributions.}
\label{fig:redist_int}
\end{figure}

We applied a traditional template fitting technique as described in
Section \ref{sec:galaxy-prop}   to acquire photometric redshift estimates for each of our
sources.  The redshift distribution for both the interacting and non-interacting
populations (Figure \ref{fig:redist_int}) is found to be comparable.  This paper probes the merger fraction over the redshift
range $0.1<z<1.2$.  The upper redshift
limit of 1.2 is due in combination to the lack of sources at
redshifts $>1.2$ (caused by the apparent magnitude limit required for
accurate visual classification), and the
fact that the 4000\AA\ break begins to move beyond our bluest filter,
reducing the accuracy of the photometric redshifts.  Ultimately, the CFHTLS Deep Catalog of Interacting galaxies
contains 1240 interacting galaxies between $0.1<z<1.2$,
$i_{vega}<21.9$, with stellar masses $>10^{9.5}M_{\odot}$.

\section{The Galaxy Interaction Fraction (GIF) at $0.1<z<1.2$}
\label{sec:galaxy-inter-fract}

Tidal tails and bridges are a result of gravitational encounters between two or
more galaxies \citep[e.g.][]{1972ApJ...178..623T}, and are nearly
foolproof signatures of a recent or ongoing galaxy interaction.
The statistics of galaxies with extended tidal features (tidal bridges and tails) is a
powerful tool to study the evolution of the galaxy interaction
fraction (GIF).  In this section we present our analysis of the merger
fraction between $0.1<z<1.2$ using the CFHTLS Deep Catalog of
Interacting Galaxies.  This catalog is the largest sample of interacting
galaxies in the literature over this redshift range.

The number of galaxies exhibiting strong tidally induced structures
like tails and bridges is a relatively simple and robust, lower limit measure of
the interaction fraction.  The CFHTLS-Deep survey is sensitive to
tidal features with surface brightnesses down to $i'\lsim$29 mag $arcsec^{-2}$.   Our
approach is simply to compare the number of galaxies with tidal
features ($N_{Int}$) to the total number of galaxies within the same
absolute magnitude or stellar mass ranges ($N_{Total}$), as
a function of redshift (see Equation \ref{eq:gif}). In the case where
a pair of galaxies are connected by a tidal bridge, each galaxy will be counted as
interacting ($N_{Int}=2$), while a galaxy with a tidal tail and no
bridge linking it to another object is counted as a single interaction ($N_{Int}=1$). 

\begin{equation}
\label{eq:gif}
Galaxy\, Interaction\, Fraction (GIF) = \frac{N_{int}}{N_{Total}}
\end{equation}

\begin{figure}[h]
  \centering
  \includegraphics[width=90mm]{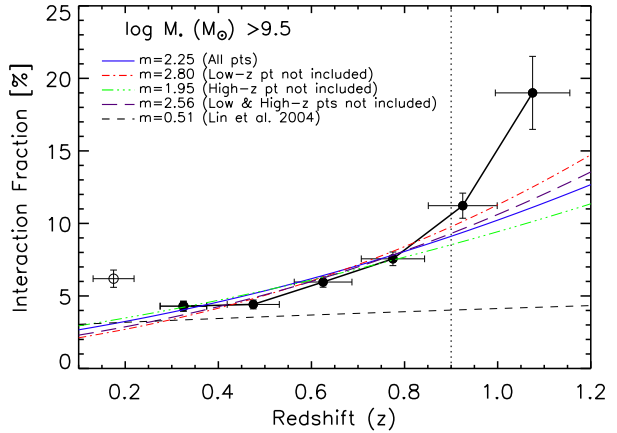}
\caption[CFHTLS-Deep: The Galaxy Interaction Fraction]{Mass limited galaxy interaction fraction as a function of redshift from the
CFHTLS-Deep Survey (circles). The colored lines represent the
best $(1+z)^{m}$ fits when various data points are included in the
analysis. When all points between $0.1\le z \le 1.15$ (with stellar
masses $\ge 10^{9.5} M_{\odot}$) are included the best fit of $(1+z)^{m}$ is
$m=2.25\pm0.24$ (blue, solid).  When the lowest redshift point is not included,
$m=2.8\pm0.25$ (red, dash-dot), only the high-$z$ point removed,
$m=1.95\pm0.2$ (green, dash-dot-dot), and
when both the low and high-$z$ points are not included $m=2.56\pm0.24$
(purple, long dash).  The black dashed line shows the result of
\citet{2004ApJ...617L...9L,2008ApJ...672..177L} ($m=0.51\pm0.28$)
who find mild to no evolution in the close pair fraction.  The error bars are derived using Poisson statistics,  while the horizontal
errors come from the uncertainty in the photometric redshift. The
vertical dotted line represents the $75\%$ completeness limit.  The open
circle notes the data point suffering from the selection bias
discussed in Section \ref{sec:potential-biases}. }
\label{fig:gif_main}
\end{figure}

In Figure \ref{fig:gif_main} we plot the fraction of galaxies with tidal tails, as a
function of redshift, assuming $i'<21.9$, and stellar masses
$>10^{9.5}M_{\odot}$. As previously described in Section \ref{sec:classification}, the interaction classes included
in this measurement are ``medium'' and ``long'' tidal tailed galaxies,
those in a close galaxy pair with a tidal bridge, and those galaxies with a double
nuclei and addition morphological evidence of an interaction such as a
tidal or bridge.  

We find that the fraction of galaxies involved in a merger rises with redshift.  Meaning,
more galaxies were undergoing a tidal interaction when the universe
was about half its current age.  At low redshift ($z\sim0.3$) the GIF
was found to be $\sim 4\%\pm$0.3, and nearly triples by $z\sim$0.95 to
$11\%\pm 0.9$ (see Table \ref{tab:gif_main}).  We characterize the evolution of the galaxy
interaction fraction by fitting a simple power-law increase with
redshift of the form
GIF=GIF$_{o}(1+z)^{m}$, where GIF$_{o}$ is the present day interaction
fraction, and $m$ is the power-law index.  When all redshift bins in
our sample are included ($0.1\le z\le1.15$) we find a best fit $m$ of
$2.25\pm 0.24$ and GIF$_{0}$ of $2.15\%\pm$0.25, weighted by the GIF
error bars in each redshift bin.  

As discussed in the
following section there may be some potential biases at the low and high
redshift ends of our analysis.  Therefore, the evolution of the GIF was
also fit using various redshift ranges as shown in Figure \ref{fig:gif_main}.  If the lowest
redshift bin ($0.1-0.25$) is excluded, a best-fit power-law index of $m=2.80\pm
0.25$ is derived.  If the highest redshift bin $1.0-1.15$ (the
least confident) is excluded $m=1.95\pm 0.25$, the lowest degree of
evolution found in our analysis.  Finally, if both the lowest and
highest redshift bins are removed, a value for $m$ of $2.56\pm 0.24$ is
found.  It is clear that even assuming the minimum value of $m$
found in this analysis, a mild to non-evolving merger fraction
($m<1.5$) with redshift is ruled out at the $\gsim4\sigma$ confidence level.

\begin{deluxetable}{cccc}
\tabletypesize{\footnotesize}
\tablecaption{\label{tab:gif_main} Inferred Galaxy Interaction
  Statistics}

\tablehead{\colhead{Redshift} &
\colhead{$N_{Int}$} &
\colhead{$N_{Total}$} &
\colhead{Interaction Fraction [$\%$]}}

\startdata

0.10-0.25 & 108 & 1745 & 6.2$\pm$0.6\\
0.25-0.40 & 160 & 3725 & 4.3$\pm$0.3\\
0.40-0.55 & 209 & 4736 & 4.4$\pm$0.3\\
0.55-0.70 & 280 & 4700 & 6.0$\pm$0.4\\
0.70-0.85 & 258 & 3413 & 7.6$\pm$0.5\\
0.85-1.00 & 168 & 1497 & 11.2$\pm$0.9\\
1.00-1.15 & 57  & 300  & 19.0$\pm$2.5\\

\enddata

\tablecomments{$N_{Int}$ specifies the number of interacting
  galaxies. $N_{Total}$ is the overall number of galaxies in the
  sample. Errors are Poisson statistics.}
\end{deluxetable}

\subsection{Potential Biases}
\label{sec:potential-biases}

In many previous merger fraction studies, any evolution measured in
the merger fraction with redshift typically depends strongly on the either the
lowest redshift ($z\lsim0.2$) data point where volume effects come
into play, or the highest most incomplete redshift bin.  Although, like many other works our
lowest and highest redshift bins suffer biases or incompleteness,
we still find evolution of the merger fraction of at least $m\sim2$ over
our most secure and complete redshift range ($0.3<z<0.9$).   

The interaction fraction for the lowest redshift bin ($0.1\ge z<0.25$) is elevated
compared to the GIF found between $0.25\le z\le 0.6$.  This is
likely related to two factors. The first is
the larger angular size of low redshift galaxies
making tidal features more easily detected. The second and likely more dominant factor is that the lower redshift
bins are more susceptible to cosmic variance as shown in Figure
\ref{fig:gif_cos} and discussed further in Section \ref{sec:gif:cosmic-variance}. 

Another potential bias to consider is that as we probe higher redshifts,
$z>0.9$, we begin to more closely observe the UV ($u^{*}$-band) which is dominated by
massive young O and B stars.  Therefore, star formation in tidal tails and bridges
may be more easily visible resulting in a higher GIF at larger redshifts.
Since the $u^{*}$-band images are not a comparable depth to the $i'$-band data, a simulation similar to that outlined in Section \ref{sec:recovery} was
run on the sub-sample of lower redshift tidal tailed galaxies but with a 1 mag enhancement 
in surface brightness.  The resultant GIF increased by $\lsim1\%$ at $z=1$.  It is
therefore unlikely that this potential bias is solely or even largely responsible for the higher GIF at $z\sim1$, especially considering that the completeness test estimates
that we are missing a much larger fraction, 15\%-30\% of tidal tailed galaxies, at $z\sim1$ due to cosmological effects.  Despite these potential biases, our analysis suggests that the evolution in the GIF is a
real physical effect, meaning their were more galaxies undergoing mergers at earlier times out to $z\sim1$.

\subsection{GIF: Cosmic Variance}
\label{sec:gif:cosmic-variance}

\begin{figure}[h]
  \centering
  \includegraphics[width=90mm]{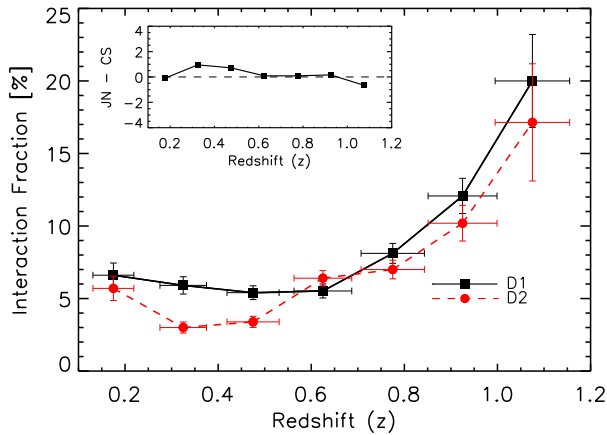}
\caption[The Galaxy Interaction Fraction: Cosmic Variance]{Galaxy interaction fraction as a function of redshift from the
CFHTLS-Deep Survey for the D1 (filled squares, black) and the D2 field
(filled circles, red).  Cosmic variance dominates the
errors of the GIF at $z<0.5$; however, at higher redshift the Poisson errors are larger. Inset panel shows the difference between
Jackknife (JN) and counting statistic (CS) errors.  }
\label{fig:gif_cos}
\end{figure}

A benefit of the CFHTLS-Deep survey is that it is spread over multiple
fields allowing for the study of how cosmic variance can affect
measurements.  In Figure \ref{fig:gif_cos}, the GIF is presented
separately for the D1 and D2 fields. Both fields show an evolving GIF with redshift, however the steepness of the evolution varies, stressing
the importance of studying multiple large fields.  The jackknife errors, representing the impact of cosmic variance, were calculated by considering
the difference in the GIF between the D1 and D2 fields (see Figure \ref{fig:gif_cos} inset).  At lower redshifts, where the volume being probed is less, we find the GIF is affected more strongly by cosmic variance, while at $z>0.6$ little variation in the GIF is seen between the fields.

\subsection{GIF: Comparison with Previous Works}
\label{sec:gif:prevwork}

A key consideration when attempting to compare results from previous
merger rate studies is the similarity of the objects being probed, i.e., comparable
luminosity, or stellar mass limits as well as the selection technique.
Numerical results of the merger fraction, using morphology should not be compared directly to a close galaxy pair fraction,
since merger selection techniques have very different observability
time-scales.  Direct comparisons, however, of the galaxy merger rate
derived with different selection techniques, can be made
since the close pair or merger fractions are normalized by the
appropriate timescale.  In Section \ref{sec:mrpwork}, we compare our merger rate with
those of previous studies.   It must be noted that the merger timescale can vary widely
depending on the initial merger parameters and remains a large
source of uncertainty.  Therefore, a comparison of merger fractions is
warranted, since they are proportional to the evolution of the merger rate.

\begin{figure}[h]
  \centering
  \includegraphics[width=90mm]{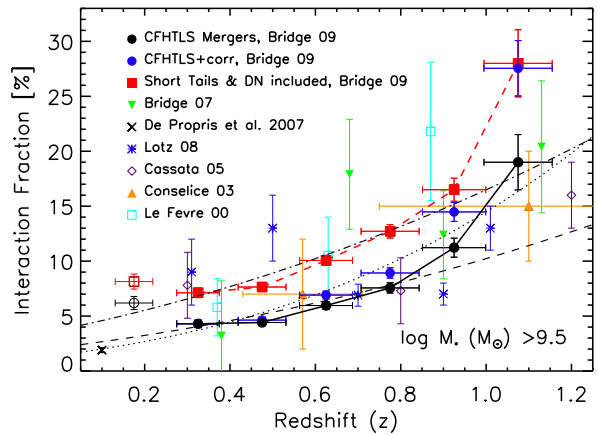}
\caption[GIF: Comparison]{Plot of the mass limited GIF as a function of redshift from the
CFHTLS-Deep Survey using the primary interacting sample
(filled circles, black) and the secondary sample (squares, red) that include
the primary merger sample and galaxies with ``short'' tidal tails, and double nuclei.  Blue circles highlight the
interaction fraction corrected for completeness.  Observations of the
merger fraction are complied from \citet[][blue
stars]{2008ApJ...672..177L}, \citet[][green, inverted
triangles]{2007ApJ...659..931B}, \citet[][black
x]{2007ApJ...666..212D}, \citet[][purple
diamonds]{2005MNRAS.357..903C}, \citet[][orange,
triangles]{2003AJ....126.1183C}, and \citet[][cyan, open
squares]{2000MNRAS.311..565L}.  The black lines outline the
best fit of form $(1+z)^{m}$ for the various samples.  $m=2.25\pm0.24$ for the
primary merger sample (dashed), while the GIF corrected for completeness (dotted
line) has a stepper fit of $m=3.31\pm0.18$.  The best fit for the
primary sample with the inclusion of probable interactions is $m=2.14\pm0.17$
(dash-dotted).  All fits include the lowest redshift points of this
study (open circle/square) which suffer from a selection bias (Section
\ref{sec:potential-biases}), strengthening $m=2.25$ as a lower limit.
It is clear that a non-evolving merger fraction is inconsistent with the CFHTLS-Deep data.}
\label{fig:gif_compare}
\end{figure}

In this section, we compare our results of the merger fraction
and its evolution with redshift with other morphologically based works.  Since we use a morphological approach in identifying
interactions a first-order comparison would be studies that also consider
the visual appearance of a galaxy whether it be qualitative or
quantitative, like those of \citet{2000MNRAS.311..565L} and
\citet{2003AJ....126.1183C} to name a couple.  Figure
\ref{fig:gif_compare} compares our work with other morphologically
based estimates of the merger fraction. The only
other study which limited their sample by stellar mass,
rather than luminosity in this redshift range is \citet{2003AJ....126.1183C}.  At
intermediate redshifts we find a merger fraction of $6\%\pm0.4\%$, in
good agreement with \citet{2003AJ....126.1183C} who find $7\%\pm5\%$,
and consistent results at $z\sim1$.
Reasonable comparisons can be made with other studies since we are also probing
similar luminosity ranges (see Section \ref{sec:gif:luminosity}).
A major obstacle that has plagued merger rate studies is small sample
sizes inspecting only a few tens to a hundred
galaxies, resulting in large
($5\%-15\%$) uncertainties making it difficult to confidently measure both the merger
fraction and its evolution with redshift.  In this study, we inspected
more than 25,000 galaxies, a factor of $\gsim10$ more than previous
morphological investigations.
We find that within the uncertainty most of the studies shown in Figure
\ref{fig:gif_compare} are consistent with our GIF measurements.

This work has presented a statistically secure lower limit of the interaction
fraction for galaxies with stellar masses $M_{*}>10^{9.5}M_{\odot}$ as a
function of redshift.  The fraction of galaxies undergoing tidal
interactions is surely higher than the rate reported here, as
this study is limited to higher surface brightness features, and only
included the most confident tidal features in its analysis.  

In Figure \ref{fig:gif_compare}, we recalculated the GIF applying the
completeness correction estimated in Section \ref{sec:recovery}.
Recall that this correction factor accounts for the number of mergers
not identified as a function of redshift due to cosmological effects.
Evolution of the galaxy merger fraction is strikingly apparent.  We
find that the best fit of (GIF$_{0},m$) is ($0.015\pm0.002, 3.31\pm0.22$).
Also shown in Figure \ref{fig:gif_compare} is the GIF with the inclusion
of less confident signatures of tidal interactions or mergers, such as
galaxies with ``short'' tidal tails (tail length less than diameter of
host), and galaxies with a double nuclei and no secondary signs of an
interaction.  
This resulted in an additional 767 interacting
galaxies being included in the analysis.  The resultant GIF is on
average $\sim 5\%$ higher than our lower limit, and its evolution with
redshift ($m=2.14\pm0.17$, compared to $m=2.25\pm0.24$) traces the lower limit GIF very well over the full
redshift range. Although these classifications were deemed to be less
confident, they are likely dominated by galaxies which are truly
merging as there is reasonable agreement with the completeness corrected GIF
at high redshift.

Overall we find that between $0.1<z<1.2$, our value of $m$ ($m=2.25$) is in good agreement with
$m=2.2\pm0.3$ found by \citet{2005MNRAS.357..903C}, $m=2.12$ reported
by \citet{2007ApJ...659..931B} and \citet{2003AJ....126.1183C}.  The merger
fraction accounting for incompleteness is in better agreement with \citet{2000MNRAS.311..565L}
which find $m=3.4\pm0.6$.  Our results do differ, however,
from \citet{2008ApJ...672..177L} ($m<1$)
 which utilized the DEEP2 survey.  There are some factors that may explain the
 discrepancy aside from the choice of luminosity versus stellar mass.
 First, the $G-M_{20}$ structural parameters have a
 significantly short observability timescale (0.2 Gyr) compared to our visual
 classification (0.8 Gyr), which could account for the lower merger
 fractions found at some redshifts.  Second, mergers were identified
 using the $Gini-M_{20}$ region calibrated by the location
 of low redshift ULIRGs on this diagram \citep{2004AJ....128..163L}.
 \citet{2007ApJS..172..329K} found that only $\sim6\%$ of visually
 classified mergers lie in the region of the Gini$-M_{20}$ used by
 \citet{2008ApJ...672..177L}.  Lotz et al.\, 2008 have also
 found through simulations that when a merger has a single nucleus the
 Gini$-M_{20}$ merger classification has a detection efficiency of
 $46\%$. Third, the mild merger evolution reported by
 \citet{2008ApJ...672..177L} is strongly tied to the two lowest
 redshift bins which suffer from the largest errors, while the merger
 fractions with the smallest uncertainty are in better agreement with
 our findings.  Finally, cosmic variance may also have an impact, as
 \citet{2008ApJ...672..177L} uses a single field on the sky.

As discussed earlier, even assuming the minimum value of $m$
found in this analysis, a mild to non-evolving merger fraction
($m<1.5$) is ruled out at the $\gsim4\sigma$ confidence level.

\subsection{GIF: Interaction Classes}

\begin{figure}
  \centering
  \includegraphics[width=90mm]{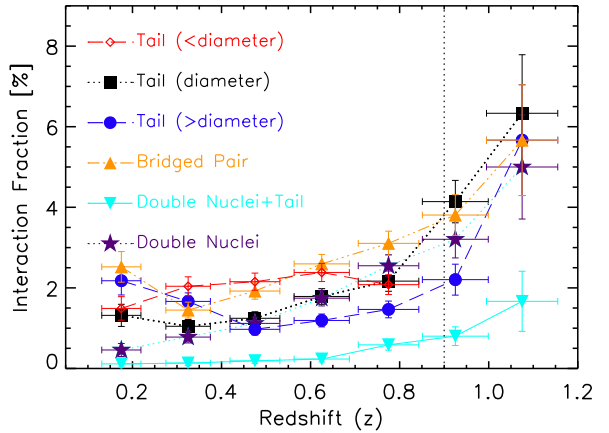}
\caption[GIF: Break-down]{Galaxy interaction
  fraction as a function of redshift for different interaction
  classes.  Short tidal tails are shown in diamonds (red), medium tails
  squares (black), long tails circles (blue), bridged close pairs
  upward triangle (orange), galaxies with a double nuclei and a tidal
  tail downward triangle (green), and double nuclei with no tail star
  (violet). The error bars are derived using Poisson statistics,  while
  the horizontal errors come from the uncertainty in the photometric redshift. The vertical dotted line represents the $75\%$ completeness limit. A stellar mass limit of $10^{9.5}(M_{\odot})$ has been imposed.}
\label{fig:gif_breakdown}
\end{figure}

The large area of the CFHTLS-Deep Survey allows us to further
explore the fraction of galaxies interacting at various stages of the
merger process and how they evolve with redshift.  Figure
\ref{fig:gif_breakdown} illustrates the GIF for the different merger
classifications. The sample
was again restricted to be brighter than $i'\le$21.9 to ensure confident
classifications and to have stellar masses $log M_{*} (M_{\odot}) \ge
9.5$ to probe similar mass galaxies at low and high redshifts.  
We find that all interacting galaxy classes show at least
some evidence of evolution with redshift.  

The specific values of the interaction
fractions for each merger class most likely vary from one another for two reasons.
First, each classification represents a snap-shot of a stage of the
merger process, and various signatures have longer durations.  For
example, a galaxy with a tidal bridge can be identified at any point
after the first encounter, and until the nuclei coalesce, while a
galaxy with a visible double nuclei and a tidal tail may be visible
for a shorter timescale as the tidal tail may have faded below the SB
limit resulting in this classification found less frequently.  A
second factor relates to the resolution of the images, as ground-based
imaging of very late stage mergers, with close double nuclei may be
blurred into a single nucleus again reducing the number of late stage mergers identified.

\subsection{Interaction Fraction: Luminosity Dependence}
\label{sec:gif:luminosity}

\begin{figure}[h]
  \centering
  \includegraphics[width=85mm]{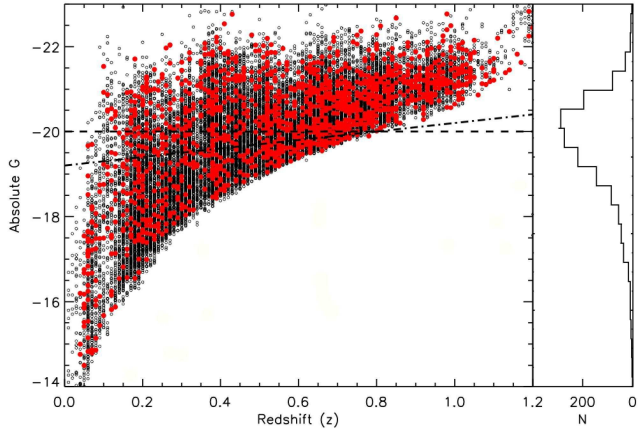}
\caption[Absolute $g'$-Band Luminosity: Interacting and Non-Interacting Galaxies]{Absolute $g'$-band luminosity as a
  function of redshift for galaxies classified as non-interacting
  (open circles, black) and those undergoing an interaction (filled
  circles, red).  The dashed lines define the lower absolute magnitude
  limits considered in the following analysis section; (dashed) no luminosity evolution, (dashed-dotted) includes
  luminosity evolution $Q=1$.  The histogram on the right shows the absolute $g'$ magnitude distribution for the interacting galaxy sample.}
\label{fig:abgdist}
\end{figure}

After estimating a lower limit for the galaxy interaction
fraction, we now explore the sensitivity of the results to various
galaxy properties.  Many previous
merger rate studies have found the merger rate to depend on optical luminosity
\citep{1997ApJ...475...29P,2000ApJ...536..153P,2003AJ....126.1183C,2004ApJ...603L..73X}.  In the following
sections, we apply various optical luminosity limits to our sample and
re-analyze the GIF to explore any possible luminosity dependence.  Figure \ref{fig:abgdist} shows the absolute $g'$-band
magnitudes ($M_{G}$) for the interacting (red) and non-interacting (black) populations in the
combined D1 and D2 fields.  It also outlines the $M_{G}$ limits imposed to aid in a
more complete comparison of galaxies at high and low redshifts with
and without luminosity evolution.

We chose a minimum ($M_{min}$) luminosity limit,
$M_{G}\le-20$, which is a few tenths of a magnitude brighter than
M$_{G}^{*}$ to balance the completeness at high redshift
with probing $M_{G}^{*}$ as closely as possible.  The sample was divided into
``bright'' ($M_{G}<-21.0$) and ``faint'' ($-21.0\le M_{G}\le-20.0$) galaxies to study the impact optical luminosity
may have on the frequency of observed galaxy interactions.  Figure \ref{fig:gif_lum} presents the GIF as a function of
redshift for the two luminosity ranges.  A clear dependence of the GIF
on $M_{G}$ is evident out to $z\sim0.7$, after which bright (blue points in Figure \ref{fig:gif_lum})  and faint (red points)
galaxies have statistically similar interaction fractions.  Within the optical luminosities outlined
above, more luminous galaxies show tidal signatures more frequently than
less luminous galaxies.  It must be considered, however, that this result may
be either entirely or in part due to the ease of identifying tidal tails in brighter galaxies, rather than a
true increase in the frequency of interactions in more luminous galaxies.  The
level of this effect is difficult to quantify and requires deeper
images of these lower luminosity galaxies to see if tidal features are evident.

Luminous galaxies, below $z<0.6$ have interaction fractions up to 2 times higher than the sample of lower luminosity galaxies.
Overall, we find that between $0.2\le z\le0.7$ the GIF for more
luminous galaxies remains fairly constant, while the GIF for fainter
galaxies shows mild evolution in this redshift range.  There is
evidence of an increasing GIF at $z>0.8$, suggesting that
interactions and merging may play a larger role in galaxy evolution at
higher redshifts.  Although this redshift bin contains $\sim$60 mergers, caution should be taken with its interpretation as it does suffer from incompleteness, particularly in the lower luminosity sample. 
At this point, we have assumed that galaxies in the past and those
locally have similar optical luminosities.  We now proceed with a
scenario that includes luminosity evolution.

\subsection{Interaction Fraction: Luminosity Evolution}
\label{subsec:gif_lum}

When a static luminosity limit is considered for a range of redshifts,
one is essentially assuming that the mass-to-light ratio ($M$/$L$) of
galaxies is the same over that redshift range. In this section,
we explore the impact that luminosity evolution may have on the GIF
measurement.   

There is considerable controversy in the literature as to
how galaxies evolve at $z<1$.  Although it is agreed
that evolution does occur
\citep{1999ApJ...518..533L,2001ApJ...563..736C}, at what magnitude
is under debate.  At the very least galaxies will evolve passively as
their stellar populations age, resulting in a gradual fading of their
optical light.  Hence, at higher redshifts when the galaxies are
younger we would expect them to have higher mean luminosities.
Detailed luminosity function studies \citep{1999ApJ...518..533L}
showed that a luminosity correction can be applied to a galaxy at redshift $z$, using the expression, $Qz$,
where $Q$ is typically taken to be equal to 1 \citep{1999ApJ...518..533L,2002ApJ...565..208P}.   

Luminosity evolution has been considered in some close
pair studies.  \citet{2002ApJ...565..208P} and
\citet{2004ApJ...617L...9L} found that $N_{c}$ (the number of
companions per galaxy) can be significantly affected by the
inclusion or exclusion of luminosity evolution.  When luminosity
evolution is corrected for, fainter galaxies at
lower redshifts are included, resulting in higher pair
statistics, and in turn less evolution of the pair
fraction with redshift.

To explore the effect luminosity evolution may have on the interaction
fraction we have adopted $Q=1$ and repeated the GIF analysis
described in the preceding section, when $Q$ was assumed to be zero.
We find a $<1\%$ difference in the GIF with and without the inclusion of luminosity evolution.  Unlike close galaxy pair's studies which typically find an increase in the number of pairs at lower redshift when luminosity evolution is applied, we find a slight decrease ($<1\%$) in the GIF , with $Q=1$.  This could merely be caused by the increased difficulty in identifying tidal tails in fainter galaxies, or perhaps there is an intrinsic dependence of the GIF on optical luminosity as implied by the ``bright'' and ``faint'' sub-samples.

\begin{figure}
  \centering
  \includegraphics[width=90mm]{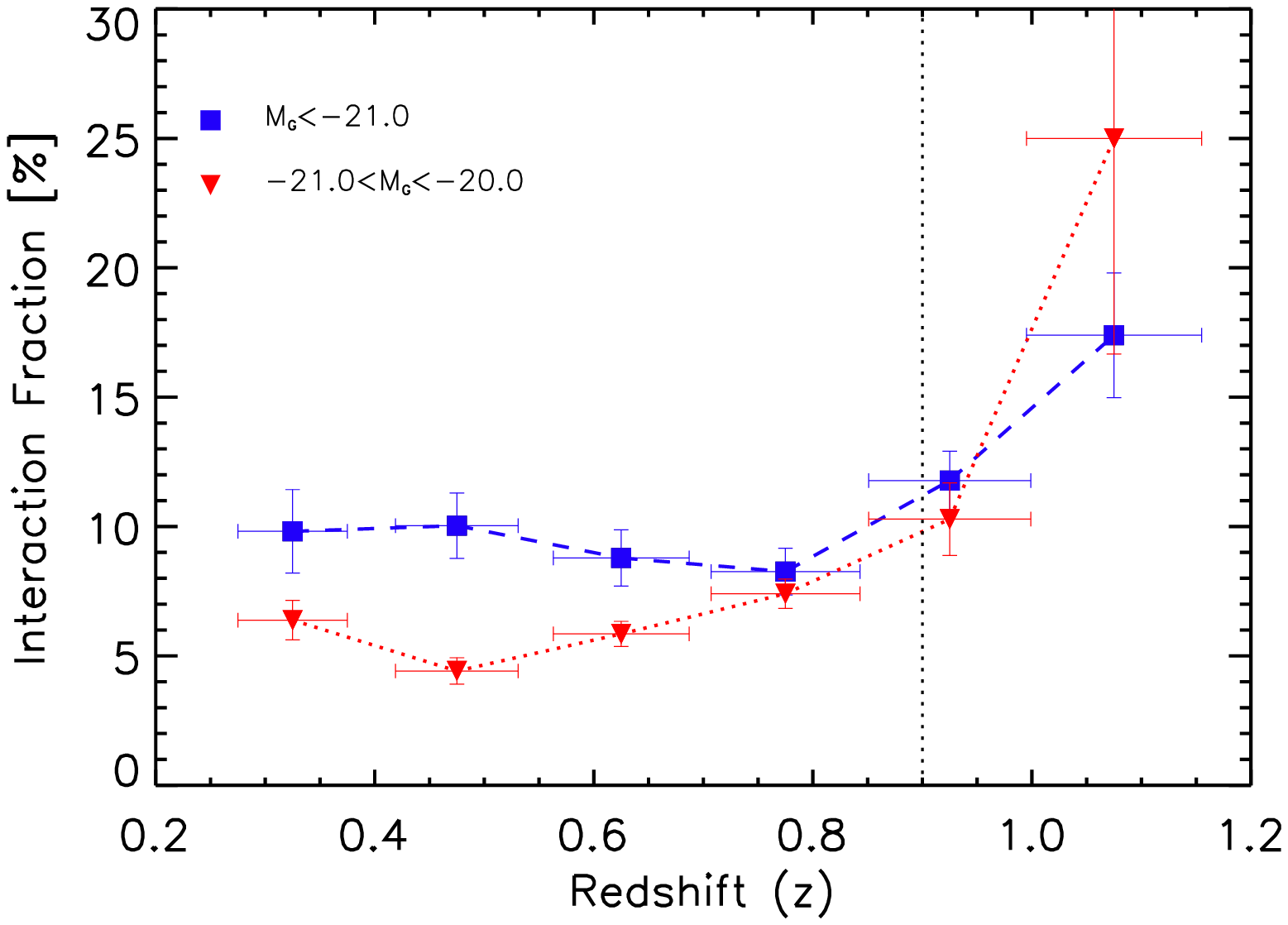}
  \includegraphics[width=90mm]{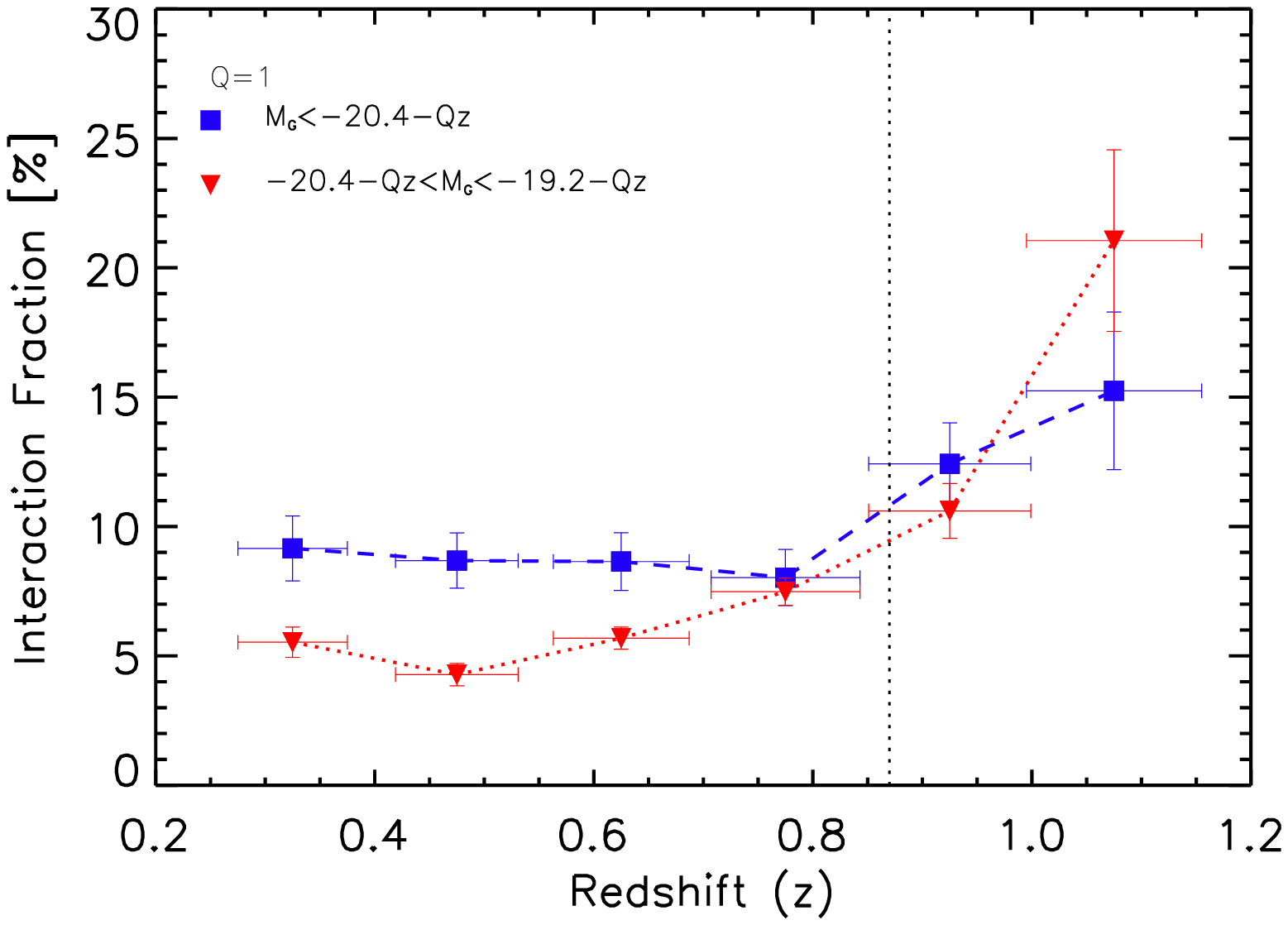}
\caption[Interaction Fraction: Optical Luminosity Dependence]{Galaxy interaction fraction as a function of redshift, assuming no
luminosity evolution, $Q=0$ (top) and $Q=1$ (bottom).  The GIF is not strongly
dependent on the assumption of luminosity evolution.  At lower redshift ($z<0.6$), more luminous galaxies (squares, blue) have a GIF up to 2 times that of lower luminosity galaxies (downward triangle, blue). The error
bars are derived using Poisson statistics,  while the horizontal
errors come from the uncertainty in the photometric redshift. The vertical dotted line represents the $75\%$ completeness limit.
}
\label{fig:gif_lum}
\end{figure}

In contrast to \citet{2004ApJ...617L...9L} who found a factor of 4
difference in $m$ with $Q=0$ to $Q=1$, we find no significant
deviation in the evolution of the GIF with or without a luminosity
evolution correction (Figure \ref{fig:gif_lum}).  This discrepancy may be an effect of the different selection techniques.  As you probe further down the luminosity function, the number of galaxies increases per unit volume, which could in turn increase the contamination from projection effects and unbound pairs.  With the tidal tail methodology, probing fainter magnitudes makes the identification of mergers more challenging.  However, the intrinsic close pair fraction or merger fraction may also be a function of luminosity.   Measuring the pair or merger fraction as a function of luminosity is highly problematic due to the numerous selection biases and contamination effects.  Disentangling these effects is challenging, which is why we have chosen to focus this study on a mass selected sample rather than a luminosity one.  

\subsection{GIF: Mass Dependence}
\label{sec:gif_mass}

Probing the merger fraction as a function of stellar mass provides
insight into how galaxies build up over time.  In hierarchical assembly, the
leading theory for structure formation in the universe, large
structures form from the merging of smaller structures.  This implies
that the most massive galaxies form latest in the history of the
universe.  Alternative theories postulate that massive galaxies can
form through rapid collapse, over short timescales in the early
universe. In this section we aim to address the merger histories of
low ($M_{*}=10^{9}-10^{10}M_{\odot}$) and high ($M_{*}>10^{10.7}M_{\odot}$) mass galaxies and
the implications this has on driving galaxy evolution.

\begin{figure}[h]
  \centering
  \includegraphics[width=90mm]{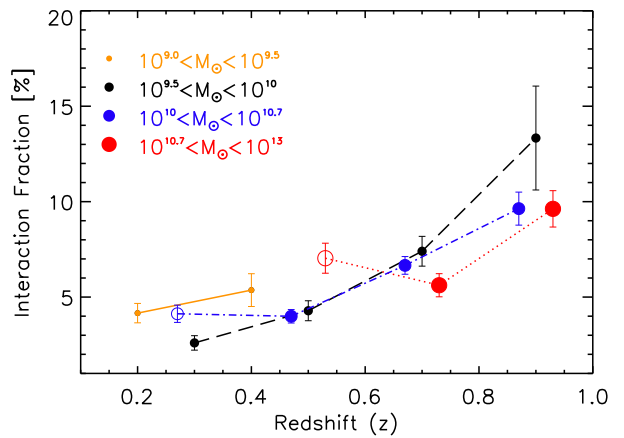}
\caption[Interaction Fraction: Stellar Mass Dependence]{Galaxy interaction fraction as a function of stellar mass and
redshift.  The data points report the GIF for four stellar
mass ranges from Log $M$=9.0 to 13, and increase in size with stellar mass.  The open data points likely suffer
from the identification bias discussed in Section
\ref{sec:galaxy-inter-fract}.  There is similar evolution of
the GIF for all mass ranges. There is an indication that lower mass
galaxies at $z<1$ interact more frequently than high-mass galaxies.
This trend is a possible mechanism for downsizing.}
\label{fig:fgif_mass}
\end{figure}

Using the five-band optical photometry in the CFHTLS, we estimated the stellar masses
of our sources through template fitting (see Section
\ref{sec:galaxy-prop} for details).  Figure \ref{fig:cfhtmass} details the stellar mass
distribution as a function of redshift for our $i'\le 21.9$ sample of
classified galaxies. Previous merger rate studies derived from observations and theoretical
models have suggested that the merger fraction not only varies with
optical luminosity as confirmed in the previous section but also
depends on stellar
mass.  It has been suggested that brighter, massive galaxies have
the highest merger fractions at $z>1$ \citep{2003AJ....126.1183C,2006ApJ...638..686C,2006ApJ...647..763M}.
In order to investigate how the GIF is affected by stellar mass, we divided our $i'\le 21.9$ sample
into four mass ranges; low masses ($10^{9.0}\le M_{*} <
10^{9.5}$ and $10^{9.5}\le M_{*} < 10^{10}$),
intermediate mass ($10^{10}\le M_{*} <10^{10.7}$), and high-mass
galaxies ($M_{*}\ge10^{10.7}$).  The interaction fraction is calculated for each stellar mass
range.  The GIF statistics are described in Table \ref{tab:tgif_mass} and plotted
in Figure \ref{fig:fgif_mass}.   All stellar mass ranges imply a
similarly evolving GIF with redshift.  It should be mentioned that the elevated merger fraction
at $z<0.6$ for massive galaxies and $z<0.4$
for intermediate mass galaxies (open circles in
Figure \ref{fig:fgif_mass}) is likely the same selection bias noted in
Section \ref{sec:potential-biases}.

\begin{deluxetable*}{lcccc}
\tabletypesize{\scriptsize}
\tablecaption{\label{tab:tgif_mass} Galaxy Interaction Statistics:
  Mass Dependence}

\tablehead{\colhead{Log Stellar Mass  ($M_{\odot}$)} &
\colhead{Redshift} &
\colhead{$N_{Total}$} & 
\colhead{$N_{Int}$} & 
\colhead{Interaction Fraction [$\%$]}}

\startdata
\tableline

$9.0\le M <9.5 $ & 0.10-0.30 & 1611 & 67 & 4.2$\pm$0.5\\
                      & 0.30-0.40 & 727 & 39 & 5.4$\pm$0.9\\
\hline
$9.5\le M <10.0$      & 0.20-0.40 & 1809 & 47 & 2.6$\pm$0.4\\
                      & 0.40-0.60 & 1564 & 67 & 4.3$\pm$0.5\\
                      & 0.60-0.80 & 1216 & 90 & 7.4$\pm$0.8\\
                      & 0.80-1.00 & 180 & 24 & 13.3$\pm$2.7\\
\hline
$10.0\le M <10.7$     & 0.20-0.40 & 2012 & 83 & 4.1$\pm$0.5\\
                      & 0.40-0.60 & 3205 & 128 & 4.0$\pm$0.4\\
                      & 0.60-0.80 & 3107 & 207 & 6.7$\pm$0.5\\
                      & 0.80-1.00 & 1287 & 124 & 9.6$\pm$0.9\\
\hline
$M \ge 10.7$       & 0.40-0.60 & 1137 & 80 & 7.0$\pm$0.5\\
                      & 0.60-0.80 & 1528 & 86 & 5.6$\pm$0.6\\
                      & 0.80-1.00 & 1060 & 102 & 9.6$\pm$1.0\\
\hline\hline\hfill
\enddata
\tablecomments{$N_{Total}$ is the number of galaxies within the specified
  stellar mass and redshift range. $N_{Int}$ the is the number of
  galaxies confidently classified as interacting within a given redshift
  and stellar mass range. The errors for the interaction fraction were
  derived using counting statistics.}
\end{deluxetable*}

A key result of Figure \ref{fig:fgif_mass} is
the indication that lower mass galaxies ($M_{*}\sim10^9-10^{10}M_{\odot}$) have
a higher merger fraction (by $\sim2\%-5\%$), compared to more massive galaxies
($M_{*}>10^{10.7}M_{\odot}$) at $z\lsim1$.  Since previous
studies at higher redshifts ($z>1.5-3$) have reported the opposite trend, finding the merger fraction of lower mass
galaxies less than that found in more massive
systems \citep{2009MNRAS.394.1956C} our result suggests (with moderate
significance) a transition of galaxy assembly between $z\sim1$ and 1.5 (i.e. galaxy assembly
downsizing). An additional implication of our findings is a potential
mechanism for interpreting cosmic downsizing.  As mergers are known to
trigger star formation \citep[][and Section \ref{sec:sfr_int} of this work]{1996ApJ...464..641M,2000ApJ...530..660B}, a higher merger fraction in lower mass
galaxies would aid in the transition of the dominant cites for star
formation moving to lower mass systems at lower redshifts.

This trend of a higher galaxy merger fraction in lower mass galaxies
at $z<1$ is in contrast with the recent close pair study in the GOODS fields of \citet{2009ApJ...697.1369B}.  A possible explanation for the higher
pair fractions (of a few percent) found in higher mass galaxies is the stronger
clustering of massive galaxies, which was not considered when
the contamination due to projection effects was estimated. Another factor to consider is the small sample size of $\lsim89$ galaxy pairs, compared to our sample of $>1100$ mergers. That being said the trend found in this work is still one of only moderate significance, stressing the need for even larger samples to better characterize the dependence of GIF on host galaxy stellar mass.

\section{Galaxy Major Merger Rate}
\label{sec:merger_rate}
The role mergers play in the formation and evolution of galaxies
is largely unknown.  The rate in which galaxies merge can affect the
mass function of galaxies, and is likely linked at some level to the
decline of the cosmic star formation rate.  In this section we discuss the galaxy major merger rate derived from the
CFHTLS-Deep Catalog of Interacting Galaxies using Equation \ref{eq:mrate} 

\begin{equation}
\label{eq:mrate}
 \Re_{mg} =GIF/T_{mg}
\end{equation}

where GIF is the galaxy interaction fraction, and $T_{mg}$ is the
assumed merger timescale.  This equation provides a measure of the
number of mergers galaxy$^{-1}$ Gyr$^{-1}$.  A primary source of uncertainty for all
merger rate studies is the merger timescale.  This observability
timescale is essentially the length of the merger process over which a specific technique
(e.g. finding close galaxy pairs or using morphology) is
able to identify the galaxy as a merger.  Estimates of merger
timescales have been derived using dynamical friction arguments
\citep{2002ApJ...565..208P} as well galaxy scale numerical
simulations tracing stellar or gas particles
\citep{2006ApJ...638..686C,2006ApJ...640..241B,2004ApJ...616..199I}. 

Recent work by \citet{2008ApJ...672..177L}, in which
a morphological analysis was performed on a suite of
$N$-body/hydrodynamical equal mass gas-rich mergers that have been
processed through a radiative transfer code have provided reasonable
estimates (0.2-0.9 Gyr), for the timescales probed by close pair
methods and quantitative morphological parameters, such
as $G-M_{20}$ and asymmetry.
\citet{2008ApJ...672..177L} shows that quantitative
morphological classifications based on $G-M_{20}$ are
sensitive during the first encounter and final merger stages for
gas-rich equal mass mergers, but do not identify many interacting
galaxies between these two stages, resulting in an observability (or
merger) timescale range of 0.2-0.6 Gyr.  Tidal tails can remain visible even after the central portion of the galaxy
exhibits a uniform or symmetric appearance, therefore the timescale our
technique is able to detect mergers is longer than that of $G-M_{20}$.

In order to estimate the merger time-scale probed by visually
identified mergers based on tidal features, we utilized $N$-body simulations of
galaxies undergoing mergers described
in \citet{2006ApJ...638..686C} which employs the models of
\citet{1996ApJ...464..641M} and \citet{2001ApJ...550...94M}.  We carefully considered the duration that a galaxy encounter would exhibit the
tidal features used in this work to identify interacting galaxies.  We
visually inspected snapshots of a simulated merger 
noting the duration in which the galaxies would be classified as ``interacting''
according to our criteria.  Ultimately, we estimate the timescale being
probed by strong visual tidal features to be 0.8 $\pm$0.2 Gyr.

\begin{figure}[h]
  \centering
  \includegraphics[width=90mm]{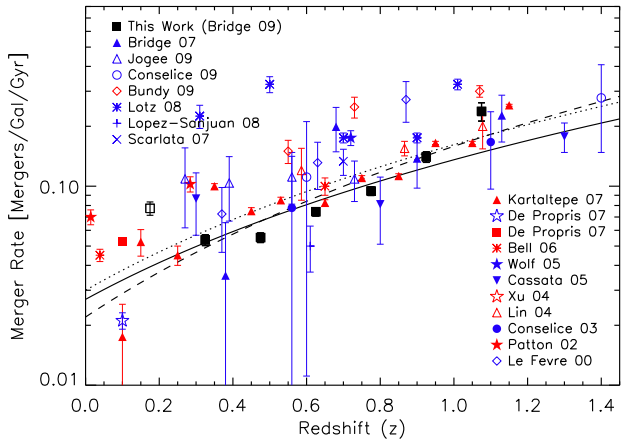}
\caption[CFHTLS Merger Rate]{Merger rate as a function of redshift
  in units of mergers galaxy$^{-1}$ Gyr$^{-1}$.  The filled black squares represent
  the merger rate derived using interacting galaxies with stellar
  masses $\ge 10^{9.5} (M_{\odot})$. The best fit to our CFHTLS data
  (solid line) of the form
  $(1+z)^{m}$ finds $m=2.33\pm0.72, $GIF(0)$=0.027\pm0.003$. Combining the
  CFHTLS-Deep data with the other works noted below results in a best
  fit with $m=2.83\pm0.29$ (dashed line) and $m=2.43$ when a $2.0\%$
  statistical error is assumed.  All fits include both the
  uncertainties of the merger fractions and merger timescale. The
  red points show the observed merger rate derived from close pair fractions as a function of
  redshift, from \citet[][filled
  stars]{2002ApJ...565..208P}, \citet[][open upward triangles]{2004ApJ...617L...9L}, \citet[][open
  star]{2004ApJ...603L..73X}, \citet[][lined
  star]{2006ApJ...652..270B}, \citet[][filled
  square]{2007ApJ...666..212D}, \citet[][filled upward
  triangles]{2007ApJS..172..320K}, and  \citet[][open
  diamonds]{2009ApJ...697.1369B}.  The merger rate derived from
  morphological studies as a function of redshift is shown in blue from
\citet[][open diamonds]{2000MNRAS.311..565L}, \citet[][filled
circle]{2003AJ....126.1183C}, \citet[][filled downward
triangles]{2005MNRAS.357..903C}, \citet[][open upward star]{2005ApJ...630..771W}, \citet[][open
star]{2007ApJ...666..212D}, \citet[][``x'']{2007ApJS..172..406S}, \citet[][filled triangle]{2007ApJ...659..931B}, \citet[][lined stars]{2008ApJ...672..177L}, \citet[][open
circle]{2008MNRAS.386..909C}, \citet[][plus sign]{2009ApJ...694..643L},   \citet[][open upward triangle]{2009ApJ...697.1971J}, and this work (filled black squares). The
assumed merger timescale for merger fractions derived using CAS
or concentration, asymmetry was $0.9\pm0.2$ Gyr from
\citet{2008MNRAS.391.1137L} and \citet{2009MNRAS.394.1956C}.  The
timescale over which mergers selected via $G-M_{20}$ is assumed to be
$0.4\pm0.2$ Gyr, close galaxy pairs $0.2\pm0.1$ Gyr
\citep{2008MNRAS.391.1137L}, and this work $0.8\pm0.2$ Gyr.}

\label{fig:merger_rate_cfht}
\end{figure}

Using Equation \ref{eq:mrate}, the merger rate for
galaxies with stellar masses $\ge 10^{9.5}M_{\odot}$ was derived and shown in
Figure \ref{fig:merger_rate_cfht} as a function of redshift.   The large uncertainty in the merger timescale
reflects the various possible merger scenarios (i.e., large mass ratios can extend the
merger timescale).

We find an average merger rate of $R_{mg}\sim$0.065 mergers gal$^{-1}$ Gyr$^{-1}$ between $0.1\le z\le0.7$, which increases to 0.24 mergers
gal$^{-1}$ Gyr$^{-1}$ at $z\sim$1.0.  The merger rate evolves with redshift as
$(1+z)^{2.25\pm0.23}$.  When both the Poisson errors in the GIF and the uncertainty in the merger timescale are included $m=2.33\pm$0.72.

Using our GIF results and the merger timescale
derived above, 
 we calculate a lower limit for
the interaction history for typical galaxies in our sample.  Equation
\ref{eq:mhistory} \citep{2006ApJ...638..686C} shows that by integrating the fraction of galaxies
undergoing an interaction divided by the merger timescale one can
obtain the number of interactions an average galaxy undergoes
between two points in redshift space.   

\begin{equation}
\label{eq:mhistory}
n_{int}=\int^{z_{1}}_{z_{2}}\frac{GIF(z)}{T_{mg}}dt = \int^{z_{1}}_{z_{2}}t_{H}(\frac{GIF_{0}}{T_{mg}})(1+z)^{m-1}\frac{dz}{E(z)}
\end{equation}

where $t_{H}$ is a Hubble time, GIF(z) is the galaxy interaction
fraction at a given redshift, GIF$_{0}$ is the GIF at $z\sim$0 , and
$E(z)=(\Omega_{M}(1+z)^{3}\Omega_{\Lambda})^{-1/2}$.
A power-law increase for the interaction rate was assumed, as it is
well fit by the data.

\begin{figure}[h]
  \centering
  \includegraphics[width=85mm]{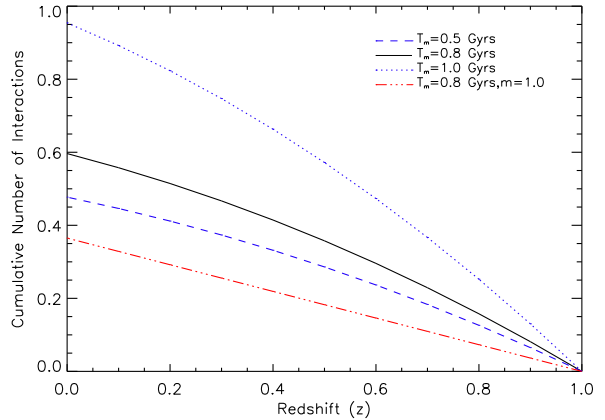}
\caption[Interaction History]{Interaction history, or the number
  of interactions an average galaxy in our sample has undergone since
  $z\sim$1 (solid, black line).  The coloured lines represent the same
  measurement but assume a different merger timescale, as stated in
  the plots legend.  The assumed evolution of the merger rate was
  $(1+z)^{2.24\pm 0.24}$.  The red (dash-dotted) line is presented for comparison
purposes assuming little evolution of the merger rate.} 
\label{fig:cnum_mergers}
\end{figure}

Based on the above equation, and assuming $m=2.25$,GIF$_{0}=0.0215$ (the
best fit to our data outlined in Section \ref{sec:galaxy-inter-fract}), we find that a galaxy with a stellar mass
$\ge 10^{9.5} M_{\odot}$ (average mass ratio $>4:1$) experiences $\sim$0.6 mergers from $z=1.0$ to
the present day (see Figure \ref{fig:cnum_mergers}).  Also shown in
Figure \ref{fig:cnum_mergers} is the number of
major mergers since $z\sim1$ derived using a range of merger
timescales.

\subsection{Merger Rate: Comparison with Previous Works}
\label{sec:mrpwork}

As discussed in Section \ref{sec:gif:prevwork} a direct comparison
between close pair fractions and morphological based merger fractions is only
robust when the measurements are converted into a merger rate through
the normalization of the appropriate observability timescale.
Figure \ref{fig:merger_rate_cfht} presents the merger rate derived
from the CFHTLS-Deep survey, and compares it with previous studies.   The assumed merger
timescales were taken from \citet{2008MNRAS.391.1137L}, where
$G-M_{20}$ has an observability timescale of $0.4\pm0.2$ Gyr, asymmetry (CAS)
$0.9\pm0.2$ Gyr, and the close pair selection, $0.2\pm0.1$ Gyr.  

We find that the merger rate evolves as $m=2.33\pm0.72$, GIF(0)$=0.027\pm0.003$. Combining the
  CFHTLS-Deep data with the literature (Figure \ref{fig:merger_rate_cfht}) results in a best
  fit with $m=2.83\pm0.29$, GIF(0)$=0.022\pm0.003$ (dashed line).  Since the fit is driven by
  data with small statistical errors we also performed a fit assuming
  a standard $2\%$ error, resulting in $m=2.43$.  All fits include both the
  uncertainties of the merger fractions and the merger timescale.
There is striking overall agreement in the literature supporting an increasing
merger rate from the local universe to $z=1.3$.  In the previous literature, it
is often suggested that the various selection techniques (e.g., close
pair versus morphology) are a source of discrepancy in terms of
measuring the evolution merger rate evolution.  Figure
\ref{fig:merger_rate_cfht} shows that studies in which the close pair
method was used to identify galaxy mergers (red data points) generally
agree with morphologically based methods (blue points).

When considering the value of $m$ derived by close pair studies our
findings are in good agreement with
 \citet[][$m=2.5\pm0.5$]{1994ApJ...429L..13B},
 \citet[][$m=2.3\pm0.7$]{2002ApJ...565..208P} and
 \citet[][$m=2.7\pm0.6$]{2000MNRAS.311..565L}. The largest close pair
 sample to date \citet{2007ApJS..172..320K} uses the 2 square degree
 COSMOS field, within which our D2 field is embedded.  The value of
 $m$ derived using close pairs in COSMOS is mildly steeper than that found in
 our full sample at $m=2.8\pm0.1$, but is consistent with our findings
 especially when we exclusively use the D2/COSMOS field (see Figure \ref{fig:gif_cos}, showing cosmic variance). It should also be noted that
 \citet{2007ApJS..172..320K} use photometric redshifts to
 select galaxies separated by $<20$ kpc, resulting in a larger
 contamination of non-merging pairs comparing to spectroscopic
 samples.  Since the accuracy of their photometric redshifts evolves as
 $0.03(1+z)$, close pair fractions at higher redshift suffer larger
 contamination which would drive the evolution of the close pair
 fraction higher, which may explain their steeper value of $m$.

 Our results do differ however from the
 close pair study of \citet{2004ApJ...617L...9L} ($m=1.08\pm0.4$) and
 morphological investigation of \citet{2008ApJ...672..177L} ($m<1$; as
 discussed in Section \ref{sec:gif:prevwork})
 both of which utilized the DEEP2 survey.  The merger rate derived using
 the close pair fractions of \citet{2004ApJ...617L...9L} is consistent
 with the merger rate values found in this work and many others;
 however, they interpret their results as evidence for a flat merger
 rate evolution.  This disparity mainly stems from the inclusion of lower redshift pair fractions from
 \citet{2000ApJ...536..153P,2002ApJ...565..208P}, which are higher
 than other pair fractions in this redshift range, when fitting the
 equation, $(1+z)^{m}$.   Another possible source of the discrepancy
 is that \citet{2004ApJ...617L...9L} find that the value of $m$ changes by a
 factor of $\sim4$ with the assumed luminosity evolution
 ($Q=0-1$), we find no strong dependence of $m$ on $Q$ (Section
 \ref{subsec:gif_lum}).  
Finally, the small sample sizes are largely unable to statistically decipher between mild and moderate merger evolution ($1<m<2$); however, the CFHTLS catalog having a sample $\sim$15 times larger is not plagued by this specific uncertainty.

\section{Star Formation Rates of Interacting Galaxies}
\label{sec:sfr_int}

Interactions and collisions can profoundly affect the evolution
of galaxies, through morphological transformations, mass accretion,
and perhaps the most brilliant, through induced star formation.  Observations of interacting
galaxies such as the Antennae \citep{1982ApJ...252..455S} and
simulations of major mergers
\citep{1996ApJ...464..641M,2006MNRAS.373.1013C} both provide
evidence that interactions can trigger violent starbursts.

\begin{figure}[h]
  \centering
  \includegraphics[width=85mm]{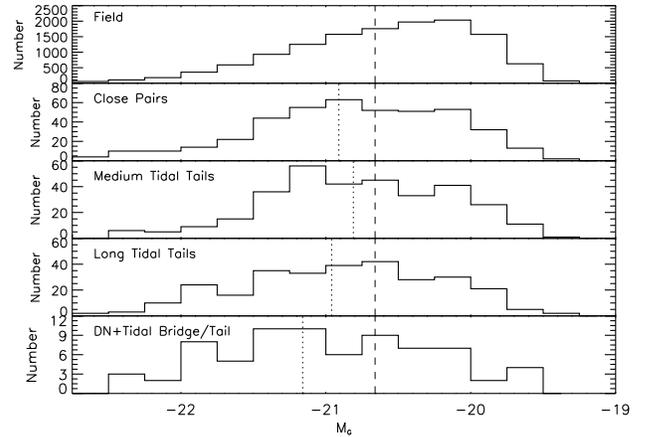}
\caption[$M_{G}$ Distribution for Interacting Galaxies]{Absolute
  $g'$-band magnitude ($M_{G}$) distribution for galaxies classified as
  interacting and the field.  The dotted line represents the average $M_{G}$ for
  each interacting class. Top: field galaxies (interacting galaxies
  removed) (second, third, fourth from top) show galaxies in a paired system with a
  tidal bridge, and galaxies with medium and long
  length tidal tails and (bottom) are galaxies with a double nuclei and
  tidal features.  The dashed line defines the average $M_{G}$ for the
  non-interacting field population. Luminosity evolution of $Q=1$ was assumed.}
\label{fig:magclass}
\end{figure}

As a first step in studying the impact mergers may have on a galaxy's
luminosity we explored the $M_{G}$ distribution for four primary
interacting galaxy types and compared them to the average
$M_{G}$ magnitude for the non-interacting field population
($M_{G}$=-20.66, Figure \ref{fig:magclass}).  The four interacting
types all show a higher average $M_{G}$ of 0.15-0.5 mag,
compared to the non-interacting population ($M_{G}$=-20.81 for medium
tidal tails, -20.91 for close pairs, -20.96 for long tails, and
-21.16 for double nuclei with tidal tails), suggesting merger
triggered star formation.

\begin{figure}[h]
  \centering
  \includegraphics[width=90mm]{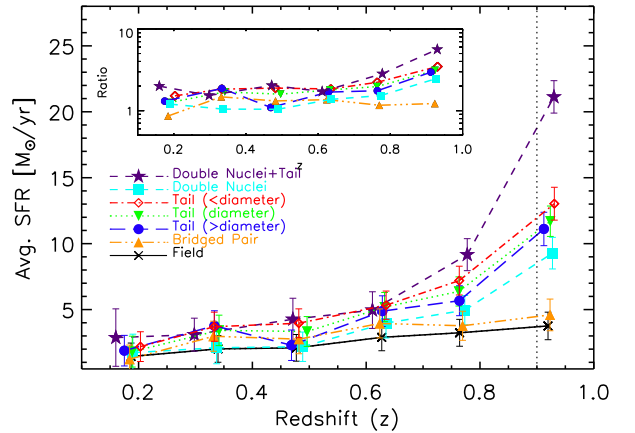}
\caption[Average Star Formation Rates: Interacting Galaxies]{Average star formation rates for various interaction classes, as
  well as the field population. Inset is the ratio of the average SFR
  of a particular class compared to the field.  The field is shown by
  the x's (black), bridged close pairs by triangle (orange), long tidal
  tails, circle (blue), intermediate tails, downward triangle (green),
  short tails, open diamond (red), double nuclei, square (cyan), and
  double nuclei with tail(s) are shown by stars (purple).  The errors
  bars are the standard deviation of the SFRs.}
\label{fig:avg_sfr}
\end{figure}

Taking a step further we used the five-band optical photometry of the CFHTLS-Deep survey and derived SFRs for each galaxy in our sample (see Section \ref{sec:galaxy-prop}). In Figure \ref{fig:avg_sfr}, we show the average
SFR for each interaction class and the non-interacting field population as a
function of redshift.  All
interaction classes from close pairs with tidal bridges (early stage
mergers) to galaxies with double nuclei and tidal tails (later stage)
exhibit enhanced SFRs a factor of 1.5-4 times that of non-interacting
galaxies.  The
level of star formation enhancement also grows with the redshift.  At
higher redshifts, one might expect the average SFR to be larger simply
because more gas is available.  We do find that the average SFR for
field galaxies (which is largely comprised of spiral
or disk dominated galaxies ($>60\%$)) increases marginally with
redshift from $\sim$1.4$M_{\odot} yr^{-1}$ at $z\sim$0.2 to
$\sim$3.2$M_{\odot} yr^{-1}$ at $z\sim$0.75.  For interacting galaxies however,
we see a factor of 3 growth in the average SFR with redshift.  This increase could be a result of these systems being
more gas-rich, allowing tidally triggered starbursts to be more significant at
higher redshifts.

Figure \ref{fig:avg_sfr} also shows that galaxies in a close pair linked by a tidal tail (an early interaction stage), 
have SFRs similar to field galaxies, while in later stage mergers (galaxies with a tidal tail or double nuclei) 
show the most enhancement, especially at $z>0.6$.  These observations agree with $N$-body
simulations which suggest that an initial starburst occurs after the
first encounter, but typically a larger burst follows at the end of
the merger sequence \citep{1996ApJ...464..641M,2005ApJ...630..705H}.

\section{Summary}

Using the deep five-band optical photometry from 2 square degrees of the
CFHTLS-Deep survey, we developed a new classification scheme to
identify galaxy mergers based on the presence of strong tidal
features.  We visually classified $\approx27,000$ galaxies, compiling the largest
catalog of interacting galaxies ($\approx1600$) in the current
literature.  With this catalog we examine the redshift evolution of
the galaxy merger fraction for galaxies with stellar masses of
$M>10^{9.5}M_{\odot}$, and the impact merging has on star formation.  
Our results can be summarized as follows.
\\

1. We find the galaxy merger fraction to be $4.3\%\pm0.3\%$ at $z\sim3$
and $19.0\%\pm2.5\%$ at $z\sim1$, implying that the frequency of galaxy interactions evolves with
redshift as $(1+z)^{2.25\pm 0.24}$.  This evolution is a lower limit.
A mild to non-evolving merger fraction
($m<1.5$) with redshift is ruled out at the $\gsim4\sigma$ confidence
level. This merger fraction study has the largest sample to date, a factor of $\gsim10$ compared to previous work.

2. The implied average merger rate between $0.2<z<0.6$ is 0.075
mergers gal$^{-1}$ Gyr$^{-1}$, which increases to 0.24 mergers gal$^{-1}$ Gyr$^{-1}$ at $z\sim1$.  Assuming $m=2.25$, a galaxy with a stellar mass
$\ge10^{9.5}M_{\odot}$ (and average mass ratio 4:1) experiences $\sim0.6$ mergers
from $z=1.0$ to the present day.

3. There is a moderately significant dependence of the merger fraction
on galaxy stellar mass.  We find that lower mass galaxies with
$M_{*}<10^{10}M_{\odot}$ are more likely to be undergoing a merger than
more massive systems ($M_{*}>10^{10.7}M_{\odot}$). This result is consistent
with expectations of galaxy assembly downsizing.

4.  A compilation of this work with previous merger rate studies,
presents an evolving merger rate, $\Re_{mg}=\Re_{0}(1+z)^{m}$, with
$m=2.83\pm0.29, \Re_{0}=0.022\pm0.003$ (errors include merger timescale uncertainty).  Also noted is the general overall agreement
between close pair and morphological selection techniques with respect
to the galaxy merger rate and its evolution with redshift. 

5. We find that major galaxy mergers have enhanced SFRs a factor of
2 higher than non-interacting galaxies. Late stage mergers
show the largest enhancement, consistent with $N$-body simulations
\citep[][to name a few]{1996ApJ...464..641M,2005ApJ...630..705H}.

Overall we have presented a consistent picture of galaxy evolution whereby galaxy interactions
occurred more frequently in the past, typically triggering bursts of star
formation. These results show that tidal tails and bridges can be a
powerful and robust tool in quantifying the interaction and merger rate
evolution of galaxies.  However it must be stressed that the merger
frequencies presented in this work are lower limits, and the fraction
of galaxies undergoing a major merger or interaction is indeed
higher, especially at larger redshifts.  

Another consideration is that
much of the star formation in the universe is enshrouded by dust.
Optical estimates of SFRs can underestimate the effect merging has on
triggering star formation in dusty systems (C. R. Bridge et al. in preparation).  It has been shown that dusty, luminous IR
galaxies (LIRGs) at $z>0.7$ are the dominant population contributing
to the comoving infrared energy density and represent $\sim70\%$ of
the star formation activity at $z\sim1$ \citep{2005ApJ...632..169L}.
So in order to compile a more complete picture of the role merging
plays in driving the CSFR it is imperative that the processes involved
in producing stars in these IR bright galaxies be understood.
We are exploring the role merging plays in LIRGs at $0.5<z<1.0$ in a companion paper (C. R. Bridge et al. in preparation).  Clearly, more work with larger samples is
needed to establish the merger rate at redshifts $>1$.  Large survey
areas with near-IR observations like those that will be provided by $HST's$
Wide Field Camera 3 will provide key data sets.  At low redshift ($<0.1<z<0.3$) the
merger rate is also poorly constrained and will benefit from future large
area, deep, multi-band surveys like that being done with the Large Synoptic Survey Telescope ( LSST).

\acknowledgments
We thank Kevin Bundy, Colin Borys and Lee Armus for their
contributions to this work and the anonymous referee for valuable comments that improved the clarity of the paper.  
 Financial support for this work was provided in part by the Natural
 Sciences and Engineering Research Council (NSERC) and an Ontario
 Graduate Scholarship in Science and Technology.   MS acknowledges
 support from the Royal Society. This paper is
 based on observations obtained with MegaPrime/MegaCam, a joint
 project of CFHT and CEA/DAPNIA, at the Canada-France-Hawaii Telescope
 (CFHT).  The authors recognize and acknowledge the very
 significant cultural role and reverence that the summit of Mauna Kea
 has always had within the indigenous Hawaiian community.  We are most
 fortunate to have the opportunity to conduct observations from this
 mountain.

\bibliographystyle{apj}
\bibliography{astro}

\end{document}